\documentclass[journal,twocolumn,table]{IEEEtran}

\usepackage{amsmath, amssymb, amsfonts, mathtools, amsthm, bbm, fdsymbol}

\usepackage{graphicx, adjustbox, subcaption, tikz, epsfig}

\usepackage{array, tabularx, booktabs, multirow, longtable, tabularray, supertabular, caption, makecell}

\usepackage{setspace, comment, float, rotating, hhline, soul, enumitem, textcase, graphicx}

\usepackage{cite, bibentry, bibunits}

\usepackage[hidelinks]{hyperref}

\usepackage[mathscr]{eucal}
\usepackage{threeparttable}
\usepackage{optidef}
\usepackage[noend]{algpseudocode} 
\usepackage[nonumberlist,nopostdot,acronym]{glossaries}
\usepackage[switch,pagewise]{lineno}
\usepackage[hmargin=1cm]{geometry}
\usepackage{nicematrix}
\usepackage{url, breakurl}

\usepackage{float}
\usepackage{orcidlink}

\usepackage{multicol}

\usepackage{longtable}
\usepackage{pdflscape} % For landscape orientation
\usepackage{array}
\usepackage{geometry}
\usepackage{adjustbox} % For rotating content
\usepackage{amsmath}
\usepackage{geometry}
\usepackage[acronym]{glossaries} 
\newacronym{3gpp}{3GPP}{3rd Generation Partnership Project}
\newacronym{5g}{5G}{Fifth Generation}
\newacronym{6g}{6G}{6th-Generation}
\newacronym{ai}{AI}{Artificial Intelligence}
\newacronym{cnn}{CNN}{Convolutional Neural Network}
\newacronym{cogsat}{CogSat}{Cognitive Satellite}
\newacronym{cr}{CR}{Cognitive Radio}
\newacronym{crsn}{CRSN}{Cognitive Radio Sensor Networks}
\newacronym{csa}{CSA}{Concurrent Spectrum Access}
\newacronym{csi}{CSI}{Channel State Information}
\newacronym{dlr}{DLR}{Deep Learning Reinforcement}
\newacronym{dsa}{DSA}{Dynamic Spectrum Allocation}
\newacronym{dsrc}{DSRC}{Dedicated Short-Range Communications}
\newacronym{fl}{FL}{Federated Learning}
\newacronym{geo}{GEO}{Geostationary Equatorial Orbit}
\newacronym{iobt}{IoBT}{Internet of Battle Things}
\newacronym{iot}{IoT}{Internet of Things}
\newacronym{isl}{ISL}{Inter-Satellite Link}
\newacronym{itu}{ITU}{International Telecommunication Union}
\newacronym{leo}{LEO}{Low Earth Orbit}
\newacronym{lstm}{LSTM}{Long Short-Term Memory}
\newacronym{madrl}{MADRL}{Multi-Agent Deep Reinforcement Learning}
\newacronym{mdp}{MDP}{Markov Decision Process}
\newacronym{meo}{MEO}{Medium Earth Orbit}
\newacronym{ml}{ML}{Machine Learning}
\newacronym{nfv}{NFV}{Network Function Virtualization}
\newacronym{ntn}{NTN}{Non-Terrestrial Networks}
\newacronym{osa}{OSA}{Opportunistic Spectrum Access}
\newacronym{pu}{PU}{Primary User}
\newacronym{qos}{QoS}{Quality of Service}
\newacronym{rl}{RL}{Reinforcement Learning}
\newacronym{rsma}{RSMA}{Rate Splitting Multiple Access}
\newacronym{satcom}{SatCom}{Satellite Communication}
\newacronym{siot}{SIoT}{Satellite Internet of Things}
\newacronym{sl}{SL}{Supervised Learning}
\newacronym{su}{SU}{Secondary User}
\newacronym{uav}{UAV}{Unmanned Aerial Vehicle}
\newacronym{usl}{USL}{Unsupervised Learning}
\newacronym{vsat}{VSAT}{Very Small Aperture Terminal}
\newacronym{wsn}{WSN}{Wireless Sensor Networks}
\newacronym{gps}{GPS}{Global Positioning Systems}
\newacronym{dsm}{DSM}{Dynamic Spectrum Management}
\newacronym{stn}{STN}{Satellite Terrestrial Network}
\newacronym{eirp}{EIRP}{Effective Isotropic Radiated Power}
\newacronym{ss}{SS}{Sectrum Sensing}
\newacronym{rtt}{RTT}{Rount Trip Time}
\newacronym{kpi}{KPI}{Key Performance Indicator}
\newacronym{ca}{CA}{Channel Availability}
\newacronym{sinr}{SINR}{Signal to Noise plus Interference Ratio}
\newacronym{snr}{SNR}{Signal to Noise Ratio}
\newacronym{inr}{INR}{Interference to Noise Ratio}
\newacronym{ipc}{IPC}{Interference Power Constraint}
\newacronym{drl}{DRL}{Deep Reinforcement Learning}
\newacronym{ddpg}{DDPG}{Deep Deterministic Policy Gradient}
\newacronym{rem}{REM}{Radio Environment Map}
\newacronym{sdn}{SDN}{Software Defined Network}
\newacronym{rnn}{RNN}{Recurrent Neural Network}
\newacronym{dl}{DL}{Deep Learning}
\newacronym{svm}{SVM}{Support Vector Machine}
\newacronym{dqn}{DQN}{Deep Q-Network}
\newacronym{dca}{DCA}{Dynamic Channel Allocation}
\newacronym{wan}{WAN}{Wide Area Network}
\newacronym{gnss}{GNSS}{Global Navigation Satellite System}
\newacronym{ppo}{PPO}{Proximal Policy Optimization}
\newacronym{vits}{VITS}{Very High Throughput Satellites}
\newacronym{llm}{LLM}{Large Language Model}
\newacronym{sdr}{SDR}{Software Defined Radio}
\newacronym{css}{CSS}{Cooperative Spectrum Sensing}
\newacronym{ieee}{IEEE}{Institute of Electrical and Electronics Engineers}
\newacronym{rssi}{RSSI}{Received Signal Strength Indicator}
\newacronym{wran}{WRAN}{Wireless Regional Area Network}
\newacronym{cstn}{CSTN}{Cognitive Satellite Terrestrial Network}
\newacronym{bs}{BS}{Base Stations}
\newacronym{noma}{NOMA}{Non-Orthogonal Multiple Access}
\newacronym{cicstn}{CI-CSTN}{Cooperative Integrated-Cognitive Satellite Terrestrial Network}
\newacronym{hts}{HTS}{High Throughput Satellite}
\newacronym{mimo}{MIMO}{Multiple Input Multiple Output}
\newacronym{fss}{FSS}{Fixed Satellite Service}
\newacronym{mss}{MSS}{Mobile Satellite Service}
\newacronym{bss}{BSS}{Broadcasting Satellite Service}
\newacronym{mifr}{MIFR}{Master International Frequency Register}
\newacronym{wrc}{WRC}{World Radiocommunication Conference}
\newacronym{ngso}{NGSO}{Non-Geostationary Satellite Orbits}
\newacronym{mtc}{MTC}{Machine-Type Communications}
\newacronym{nbiot}{NB-IoT}{Narrowband Internet of Things}
\newacronym{etsi}{ETSI}{European Telecommunications Standards Institute}
\newacronym{bsm}{BSM}{Broadband Satellite Multimedia}
\newacronym{haps}{HAPS}{High-Altitude Platform Stations}
\newacronym{embb}{eMBB}{Enhanced Mobile Broadband}
\newacronym{mmtc}{mMTC}{Massive Machine Type Communications}
\newacronym{urllc}{URLLC}{Ultra-Reliable Low-Latency Communications}
\newacronym{ebu}{EBU}{European Broadcasting Union}
\newacronym{dth}{DTH}{Direct-to-Home}
\newacronym{fcc}{FCC}{Federal Communications Commission}
\newacronym{acma}{ACMA}{Australian Communications and Media Authority}
\newacronym{cept}{CEPT}{European Conference of Postal and Telecommunications Administrations}
\newacronym{citel}{CITEL}{Inter-American Telecommunication Commission}
\newacronym{apt}{APT}{Asia-Pacific Telecommunity}
\newacronym{aws}{AWS}{Amazon Web Services}
\newacronym{esa}{ESA}{European Space Agency}
\newacronym{qpsk}{QPSK}{Quadrature Phase Shift Keying}
\newacronym{qam}{QAM}{Quadrature Amplitude Modulation}
\newacronym{tvws}{TVWS}{Television White Spaces}
\newacronym{darpa}{DARPA}{Defense Advanced Research Projects Agency}
\newacronym{iss}{ISS}{International Space Station}
\newacronym{scan}{SCaN}{Space Communications and Navigation}
\newacronym{knn}{KNN}{K-Nearest Neighbours}
\newacronym{pca}{PCA}{Principal Component Analysis}
\newacronym{gpt}{GPT}{Generative Pre-trained Transformer}
\newacronym{gan}{GAN}{Generative Adversarial Network}
\newacronym{fpga}{FPGA}{Field-Programmable Gate Array}
\newacronym{gpp}{GPP}{General-Purpose Processor}
\newacronym{vnf}{VNF}{Virtual Network Function}
\newacronym{mec}{MEC}{Multi Access Edge Computing}
\newacronym{usa}{USA}{United States of America}
\newacronym{uk}{UK}{United Kingdom}
\newacronym{vae}{VAE}{Variational Autoencoder}
\newacronym{los}{LoS}{Line of Sight}

\makeglossaries

\begin{document}

\title{Intelligent Spectrum Management in Satellite Communications}

\author{
    Rakshitha De Silva \orcidlink{0000-0002-7194-7619}, 
    Shiva Raj Pokhrel \orcidlink{0000-0001-5819-765X}, 
    Jonathan Kua \orcidlink{0000-0001-9699-9418} and
    Sithamparanathan Kandeepan \orcidlink{0000-0002-9388-9173} 
    \thanks{This work is supported by SmartSat CRC, whose activities are funded by the Australian Government’s CRC Program.\\
    R.~De~Silva, S.~R.~Pokhrel and J.~Kua are with the IoT \& Software Engineering Research Lab, School of Information Technology, Deakin University, Geelong, VIC 3125, Australia (e-mail: \href{mailto:rakshitha.desilva@deakin.edu.au}{rakshitha.desilva@deakin.edu.au}; \href{mailto:shiva.pokhrel@deakin.edu.au}{shiva.pokhrel@deakin.edu.au}; \href{mailto:jonathan.kua@deakin.edu.au}{jonathan.kua@deakin.edu.au}).\\
    S. Kandeepan is with the Electronic and Telecommunication Engineering at RMIT University, Melbourne, Australia (e-mail: \href{mailto:kandeepan@rmit.edu.au}{kandeepan@rmit.edu.au}). \\
    }
}

\maketitle

\begin{abstract}
\gls{satcom} networks represent a fundamental pillar in modern global connectivity, facilitating reliable service and extensive coverage across a plethora of applications. 
% As \gls{geo}, \gls{meo}, and \gls{leo} satellites possess unique advantages enabling specific use cases, 
The expanding demand for high-bandwidth services and the proliferation of mega satellite constellations highlight the limitations of traditional exclusive satellite spectrum allocation approaches.
\gls{cr} leading to \gls{cogsat} networks through \gls{dsm}, which enables the dynamic adaptability of radio equipment to environmental conditions for optimal performance, presents a promising solution for the emerging spectrum scarcity. 
In this survey, we explore the adaptation of intelligent \gls{dsm} methodologies to \gls{satcom}, leveraging satellite network integrations.
We discuss contributions and hurdles in regulations and standardizations in realizing intelligent \gls{dsm} in \gls{satcom}, and deep dive into \gls{dsm} techniques, which enable \gls{cogsat} networks.
Furthermore, we extensively evaluate and categorize state-of-the-art \gls{ai}/\gls{ml} methods leveraged for \gls{dsm} while exploring operational resilience and robustness of such integrations.  
In addition, performance evaluation metrics critical for adaptive resource management and system optimization in \gls{cogsat} networks are thoroughly investigated. This survey also identifies open challenges and outlines future research directions in regulatory frameworks, network architectures, and intelligent spectrum management, paving the way for sustainable and scalable \gls{satcom} networks for enhanced global connectivity.
\end{abstract}


\section{Introduction}

\IEEEPARstart{S}{atellite} networks have emerged as a cornerstone of global communication, providing extensive coverage and reliability with enhanced bandwidth across numerous applications. 
Operating in three primary orbits—\gls{leo}, \gls{meo}, and \gls{geo}, each satellite system offers distinct characteristics tailored for specific use cases. 
Mega \gls{leo} constellations, such as OneWeb, SpaceX's Starlink and Amazon's Kuiper, are transforming global connectivity by delivering low-latency, high-speed internet to remote and underserved regions~\cite{pachler2021updated}. 
Meanwhile, \gls{meo} satellites play a vital role in \gls{gps}, offering accurate navigation and timing services while playing a prominent role in civilian and military communication. 
\gls{geo} satellites operate in a fixed position relative to the Earth, remain essential for broadcasting, weather monitoring, and latency-resilient communication over vast areas.

% Multi-orbital satellite integrations leverage the advantages of different orbital regimes, combining the wide coverage of \gls{geo} with the low-latency, high-throughput capabilities of non-\gls{geo} satellites. 
% Satellite-terrestrial network integrations are an emerging topic to facilitate seamless user connectivity, breaking the barriers of conventional networks. 
% Such integrations enhance network resilience, bandwidth efficiency, and service continuity, making \gls{satcom} networks highly adaptable to a wide range of applications, from global broadband to disaster response and defence.
The growing number of satellites and rising user demand highlight the need for efficient spectrum utilization in \acrfull{satcom} networks, as limited communication spectrum remains a primary barrier for new \gls{satcom} operators to enter the market. 
This scarcity also intensifies competition for spectrum access, inadvertently contributing to higher service costs for users.
\acrfull{dsm} through \acrfull{cogsat}, the integration of \acrfull{cr} to \gls{satcom} networks emerges as a solution to this problem, leveraging intelligent and advanced solutions to improve the spectrum utilization. 
Traditional spectrum management methods, often static, constrained and pre-defined, are no longer sufficient to cope with the distributed, heterogeneous, and congested nature of modern satellite networks. 
In recent years, \acrfull{ai} and \acrfull{ml} have emerged as powerful enablers for addressing these challenges by enabling data-driven adaptive decision-making. 
Through real-time traffic patterns, spectrum occupancy, and environmental conditions, \gls{ai}/\gls{ml} techniques can optimize spectrum utilization, mitigate interference, and support \gls{cogsat} systems across multiple orbital and terrestrial networks \cite{liang2020dynamic, zhang2022spectrum}.

\subsection{Motivation}

\begin{figure*}[!h]
    \centering
    \includegraphics[width=\textwidth]{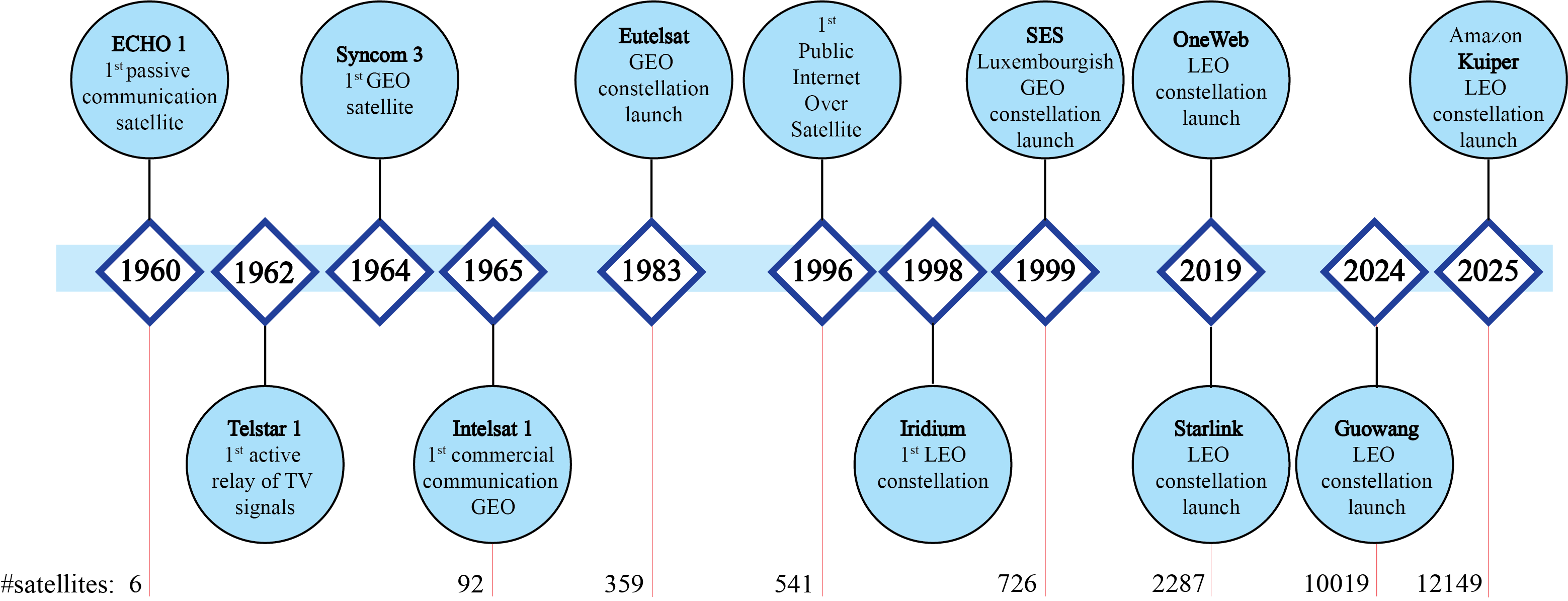}
    \caption{Exponential growth of the satellite communication industry.}
    \label{sat_timeline}
    \vspace{-5mm}
\end{figure*}

Technological advancements, including reusable rockets and ride-share programs, have accelerated the growth of modern satellite networks, spurred by the increasing demand for high-capacity broadband and resilient communication systems. 
The exponential growth of the \gls{satcom} industry is illustrated in Fig. \ref{sat_timeline}. 
Reports indicate there are 12,149 active satellites are orbiting the Earth in mid-2025, and this number is growing exponentially, while communication satellites account for 79\% of satellites in space~\cite{bryce}.
The mega \gls{leo} satellite constellation boom started with OneWeb and Starlink launches has skyrocketed the number of active satellites in orbit and with Guowang and Amazon Kuiper, it is expected to multiply. 
With this augmentation, the need for optimizing satellite operations under environmental and operational constraints is paramount. 
Specifically, the \gls{satcom} spectrum should be utilized optimally as it is one of the most sought-after finite resources in this growing industry.
However, the existing exclusive licensing approach has led to an artificial scarcity of the radio communication spectrum. 
This growing mismatch between regulatory rigidity and operational complexity underscores the critical need for dynamic and intelligent spectrum management frameworks that can be integrated into \gls{satcom} networks.

% \textcolor{blue}{need to dsm and cogsat}
To address these challenges, the concept of \gls{cogsat} networks has emerged as a transformative approach, integrating \gls{cr} into the \gls{satcom} environment, enabling intelligent spectrum access and autonomous resource control. 
% \gls{cr} enables radios to dynamically alter transmission strategies such as frequency, bandwidth, modulation, and power to optimize performance. 
\gls{cr} enables sharing frequency, space, time, and power information capacity dimensions through dynamically altering radio transmission strategies.
Through \gls{cr}, \gls{dsm} enables to leverage of unutilized or underutilized spectrum, improving spectral efficiency.
\gls{cogsat} systems extend these into the unique constraints of space, incorporating native \gls{satcom} parameters such as orbital dynamics, multi-orbital network coordination, and delay-sensitive network configurations \cite{kodheli2020satellite}. 
% CR introduced by Joseph Mitola in 1999, represents a paradigm shift in wireless communication through intelligent adaptation to the surrounding environment. 
% Given the rapid expansion of \gls{satcom} networks and the finite availability of dedicated frequency spectrum as highlighted above, extending CR principles to satellite networks is crucial for sustainable growth and seamless global connectivity.
% \gls{cogsat} networks, the adaptation of CR concepts into satellite networks, are envisioned to autonomously and intelligently manage \gls{satcom} resources and transmission properties \cite{SMT2}. 
% Unlike traditional satellite systems, \gls{cogsat} networks are expected to leverage advanced cognitive mechanisms to autonomously optimize spectrum utilization, enhance throughput, and maintain communication reliability under highly congested, contested, or adversarial conditions \cite{7336495}. 
% This paradigm shift is particularly critical given the rapid proliferation of satellite deployments, which has led to increasingly saturated orbital and spectral domains. 
Through the integration of cognitive functionalities into the satellite architecture, \gls{cogsat} networks aim to significantly elevate the resilience and efficiency of space-based communication systems, thereby addressing the pressing demands of modern commercial and defence \gls{satcom} applications \cite{gupta2015cognitive}.

% As satellite networks transition toward intelligent, autonomous, and densely populated architectures, the regulatory and standardization landscape must evolve in parallel to support these advancements. 
Traditional spectrum governance are characterized by rigid, long-term, and exclusive allocations, which poses a significant bottleneck to the adoption of dynamic and cognitive spectrum access strategies to \gls{satcom}. 
Without a coordinated regulatory framework that embraces flexibility, interoperability, and real-time spectrum sharing, the benefits of intelligent spectrum management cannot be fully realized. 
Standardization efforts led by bodies such as the \gls{itu}, \gls{3gpp}, \gls{ieee} and \gls{etsi} are therefore essential to establish harmonized protocols, interoperability requirements, and policy guidelines that accommodate cognitive functionalities, orbital diversity, and cross-network coordination. 
On the other hand, the performance evaluation metrics for these dynamic and intelligent spectrum management approaches are yet to be fully explored. 
Therefore, the discussions on regulations and standardizations alongside \gls{dsm} performance indicators is not only timely but also fundamental to ensuring that \gls{cogsat} networks can operate efficiently, securely, and equitably within an increasingly congested and contested spectrum environment~\cite{centenaro2021survey}.

% \textcolor{blue}{need of ai and ml}
\gls{ai} and \gls{ml}, known for their proficiency in pattern recognition and decision-making, are key enablers of this vision. 
In \gls{cr}, \gls{ml} techniques are already applied to optimize resource allocation, interference management, and spectrum access~\cite{al2023artificial}. 
In the context of \gls{satcom} networks, \gls{ml} can address the challenges of atmospheric effects, propagation delays, and dynamic interference levels, facilitating the realization of \gls{cogsat} systems~\cite{fontanesi2025artificial}. 
\gls{sdr} brings this flexibility to radio networks through allowing the control of radio parameters through software programs. 
On the other hand, \gls{sdn}, the paradigm of decoupling control and data planes enables new avenues towards flexible networking. 
The combination of \gls{sdn} and \gls{sdr} is a key enabler of \gls{cr} and associated \gls{dsm} techniques, and they are widely utilized in modern network integrations. 
\gls{nfv} enables network functions such as routers and firewalls to run as \glspl{vnf}, allowing them to operate beyond vendor proprietary boundaries. 
These technologies are widely leveraged in terrestrial and non-terrestrial network deployments, as they provide unprecedented flexibility. 
\gls{cogsat} networks empowered through AI and \gls{ml} and upheld through \gls{sdn}, \gls{nfv}, and privacy-preserving techniques like blockchain, have the potential to meet the growing demands of data-intensive applications in future satellite networks. 

% These cognitive capabilities, when coupled with \gls{ai} and \gls{ml} in satellite systems empowered with \gls{sdn}, \gls{nfv}, and edge computing, allow satellites to operate as intelligent agents within a dynamic collaborative network fabric. 
% As such, \gls{cogsat} networks are poised not only to alleviate spectrum scarcity but also to redefine the resilience, scalability, and efficiency of next-generation \gls{satcom}s, ensuring sustainable growth.

\begin{table*}[!h]
\caption{{Related literature surveys and contributions}}
\centering
\resizebox{0.8\textwidth}{!}{
    \begin{tabular}{|p{1.4cm}|p{1.8cm}|p{1.5cm}|p{1.2cm}|p{1.2cm}|p{1.5cm}|p{8.6cm}|}
    \hline
    Reference & Regulations \& standardizations & Network Architectures & \gls{cr} \& \gls{dsm} & \gls{ai} \& \gls{ml} & Performance Metrics & Key Contributions \\ \hline
    \cite{fontanesi2025artificial} & N  & L & M & H & N & Discuss the limits and constraints of AI \& \gls{ml} integration for \gls{satcom} onboard operations, along with different use cases and evaluate possible hardware solutions. \\ \hline
    \cite{liu2020review} & N  & L & H & N & L & Elaborate on CR and application scenarios of \gls{cogsat} communication. Summarises work on \gls{ss}, spectrum allocation and power control. \\ \hline
    \cite{jasim2021survey} & H  & M & H & H & N & Spectrum management regulations, architectures, approaches, and tools for Unmanned Aerial Vehicles (UAVs) communication.  \\ \hline
    \cite{kodheli2020satellite} & H  & H & L & L & N & Technological analyses on \gls{satcom}. Access control and networking challenges of satellite networks with testbed outcomes.  \\ \hline
    \cite{zhang2022spectrum} & M  & M & H & N & N & Spectrum sharing in aerial/space networks, with techniques, spectrum utilization rules and associated key technologies.   \\ \hline
    \cite{heydarishahreza2024spectrum} & N & H & M & M & M & Explore \gls{leo}-Terrestrial network integration in the context of interference in different network deployment scenarios.  \\ \hline
    \cite{ortiz2022machine} & N  & N & H & H & L & \gls{ml} for radio resource management in \gls{geo} satellites\\ \hline
    \cite{al2023artificial} & N  & M & L & H & N & AI techniques for integrated terrestrial-massive satellite networks.  \\ \hline
    Ours & H  & H & H & H & H & Explore regulation and standardization with existing \gls{dsm} techniques along the line of AI \& \gls{ml} for \gls{satcom} networks. Discuss the network architectures that enable \gls{cogsat} in \gls{satcom} networks and detail KPIs for such networks.  \\ \hline
    \multicolumn{7}{c}{ N - No impact, L - Low impact, M - Medium impact, H - High impact }
    \end{tabular}
    }
    \label{survey_papers}
    \vspace{-5mm}
\end{table*}

Therefore, it is paramount to explore how such technologies can be leveraged to enable intelligent \gls{dsm} in \gls{satcom} networks to improve spectral efficiency. 
Table \ref{survey_papers} presents an overview of the existing literature surveys on \gls{ai} \& \gls{ml} integration, \gls{dsm}, regulation, and standardization of satellite and high altitude networks with their key contributions.
The majority of work on intelligent spectrum management along \gls{cr} methodologies focuses on terrestrial communication networks; while some survey work targets specific directions of \gls{cr}, such as \gls{ss} and \glspl{rem}. 
The work discusses \gls{ai} \& \gls{ml} approaches for satellite networks, generally deep dive into the limitations and challenges of integrating them in \gls{satcom} networks without discussing the intelligent \gls{dsm} and associated challenges. Contributions and gaps in the regulation and standardization of intelligent spectrum management are rarely discussed alongside performance metrics for \gls{satcom} in the literature. 

\subsection{Contributions}

A summary of the main contributions of this paper is as follows: 
\begin{itemize}
    \item We extensively discuss the enablers of intelligent \gls{dsm} in \gls{satcom} along the lines of satellite network integrations, \gls{cr}, \gls{ai}/\gls{ml}, \gls{sdn}, \gls{nfv} and edge computing. 
    \item We explore the existing regulatory and standardization bodies on \gls{satcom} networks in the context of spectrum management and network integrations, with their contributions towards the advancement of the \gls{satcom} industry.
    \item We further explore \gls{osa}, \gls{csa}, \gls{ss}, and database techniques for \gls{satcom} along with an extensive evaluation of satellite network architectures leading to \gls{cogsat} networks. In addition, we categorize literature on \gls{dsm} for \gls{satcom} based on core functionalities and \gls{dsm} techniques.
    \item We investigate \gls{ai} and \gls{ml} methods leveraged in \gls{ss}, spectrum allocation, interference mitigation and resource management. We further discuss \gls{ml} model training and operational resilience, while extensively categorizing the state-of-the-art \gls{ml} methods on satellite spectrum management.
    \item Accurate performance evaluation methods are essential for adaptive resource management and overall system performance optimization in \gls{cogsat} networks. We focus on such metrics and discusses their evaluation criteria. 
    \item Finally, we highlight the challenges and future directions in regulatory, architectural, and \gls{ml} implementations in the context of realizing \gls{cogsat} systems toward sustainable and scalable \gls{satcom} networks for global connectivity. 
\end{itemize}

\subsection{Paper Organization}
The remainder of this article is organized as follows. 
% Section \ref{on_sat} discusses the modern \gls{satcom} networks and realized satellite network integrations. 
Section \ref{sec:enablers} elaborates on the key enablers of intelligent spectrum management in \gls{satcom}. 
Section \ref{sec: regulations} discusses the regulations and standardizations established by the prominent authorities on satellite spectrum management. 
Section \ref{sec: dsm for cogsat} focuses on \gls{dsm} for \gls{cogsat}, and Section \ref{sec: ML in Sat} discusses the state-of-the-art \gls{ml} techniques proposed in the literature for satellite spectrum management and categorizes them based on their primary focus areas. 
Section \ref{sec:performance} details the performance evaluation metrics for intelligent spectrum management in \gls{satcom} and Section \ref{sec:challanges} extensively discusses the challenges in realizing \gls{dsm} through \gls{cogsat} networks within the existing framework. Finally, Section \ref{conclusion} concludes the paper. 
% An overview of this survey paper is illustrated in Fig. \ref{fig:overview}. 

\section{Enablers of Intelligent Spectrum Management in Satellite Communications}
\label{sec:enablers}

\subsection{Satellite Network Integration}

\subsubsection{Intra-Satellite Network Integration}
\begin{figure}[H]
    \centering
    \includegraphics[width = 0.8\columnwidth]{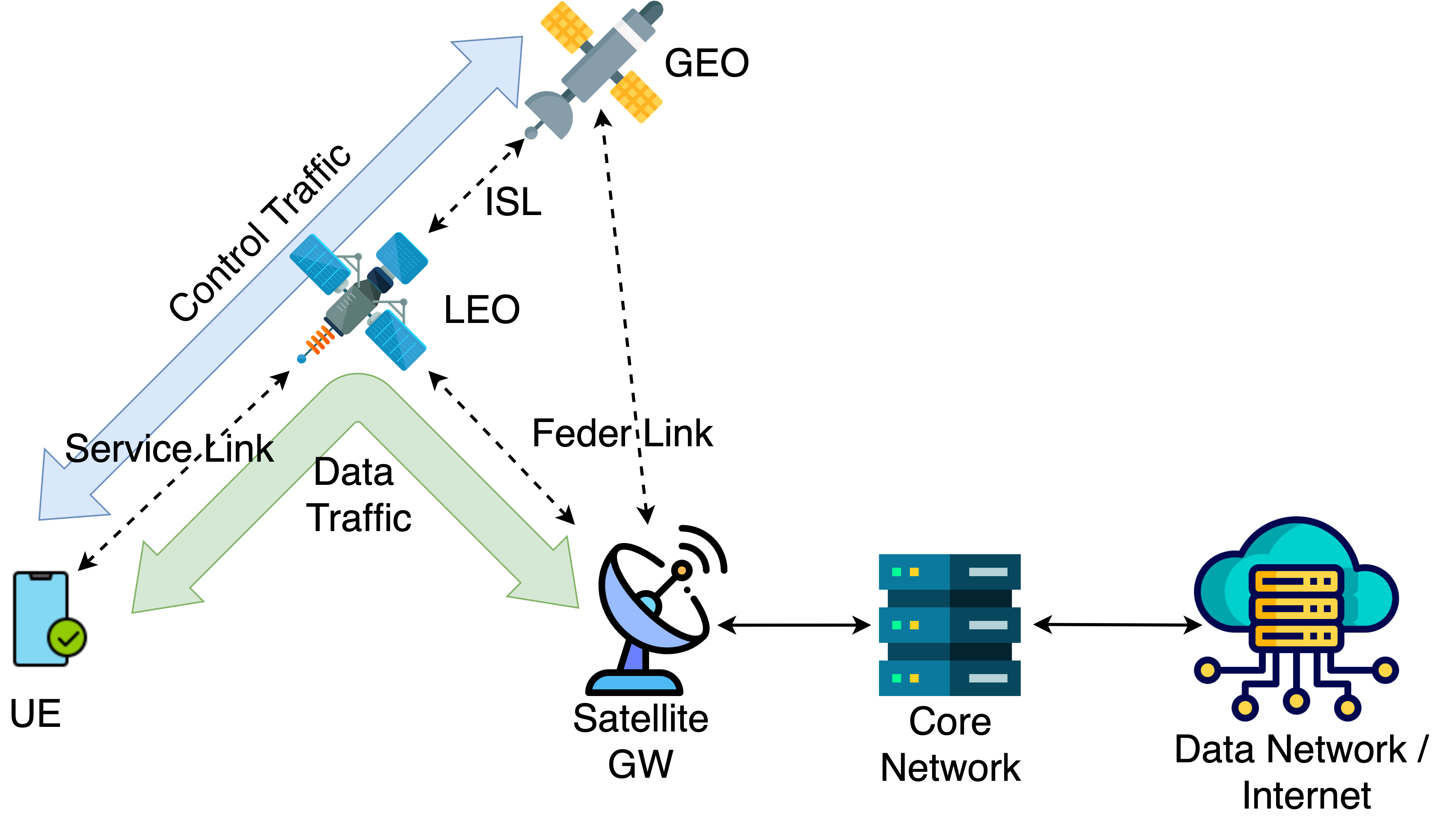}
    \caption{LEO and GEO integration architecture \cite{10211619}.}
    \label{fig:leo_geo_inte}
\end{figure}

\gls{geo}, \gls{meo}, and \gls{leo} satellite networks leverage the strengths of different orbits to provide enhanced connectivity. 
Due to the unique capabilities of these satellite networks, network operators are leaning toward utilizing the \gls{satcom} spectrum harmoniously to enhance service delivery and widen the subscriber base with hybrid satellite networks. 
Although the concept of integrating different satellite networks has been discussed, connectivity is typically established through gateways, and networks are operated separately. 
Therefore, a higher level of integration between these satellite networks is required to leverage the full scale of capabilities.
Fig. \ref{fig:leo_geo_inte} illustrates a \gls{geo}-\gls{leo} integrated network architecture where control and data traffic are separated, which leverages \gls{sdn} and \gls{nfv} technologies \cite{10211619}. 
A similar network setup where \gls{geo} satellites operate as the control layer is discussed and evaluated in \cite{afzali2023integrating}.  

In addition to the theoretical analysis of inter-satellite network integrations, Intelsat, a \gls{geo} satellite operator, and OneWeb demonstrated broadband connectivity through an integrated \gls{geo} and \gls{leo} network for the U.S. Army. 
The demonstration exemplified a throughput increase in twofold as the users connect to both \gls{leo} and \gls{geo} networks simultaneously and route blended traffic, leveraging both satellite networks~\cite{intelsat_oneweb}. 
Eutelsat and OneWeb combined their resources in 2023 to become the world's first \gls{leo}-\gls{geo} integrated satellite operator. 
They provide back-haul and corporate network connectivity for fixed users and mobile connectivity for maritime and in-flight users \cite{eutelsat}. 
In September 2023, \gls{meo} satellite operator SES announced broadband connectivity, integrating its O3b mPOWER \gls{meo} satellites and Starlink's \glspl{leo} for cruise ship passengers. 
The service is expected to facilitate data and voice over the internet for a guaranteed throughput of up to 3 Gbps per ship located anywhere in the world \cite{ses}.

\subsubsection{Satellite-Terrestrial Network Integration}
\gls{5g} and beyond cellular networks are expected to have seamless integration with satellite networks, thus taking another step toward solving the challenging problem of global connectivity \cite{latva2019key}. 
Coverage through satellite networks to remote areas is identified as a cost-effective alternative for expensive terrestrial network expansions \cite{Telstra_5gNTN}. 
On the other hand, satellite networks provide a unique advantage for maritime and aviation communication requirements.  
However, even though satellite networks are equipped to facilitate remote connectivity requirements, satellite and terrestrial cellular networks operate as separate entities connected through network gateways. 
Therefore, a tight integration between the two networks is required to facilitate low latency and high bandwidth requirements that modern user applications demand. 
The following are concrete examples of \gls{5g} terrestrial and satellite network integrations to deliver connectivity over the hybrid network setup. 

\begin{enumerate}
    \item Ericsson, Qualcomm, and Thales Alenia Space: Collaboration of these three companies successfully initiated the world's first publicly announced integration of \gls{5g}  \gls{ntn}-based networks set up on \gls{3gpp} standards and tested a \gls{5g} standard call through \gls{leo} satellite channels.
    Their test-bed has accounted for inherited challenges such as delay and Doppler effects while ensuring seamless satellite handovers \cite{Ericsson_5gNTN}.
    \item Vodafone and AST SpaceMobile: The two companies collaboratively delivered the world's first space-based \gls{5g} voice call using an unmodified regular \gls{5g} enabled smartphone. The direct-to-device test call was made from Hawaii to José Guevara through AST SpaceMobile’s BlueWalker 3 \gls{leo} \cite{Vodafone_5gNTN}.  
    \item Sateliot and \gls{aws}: A \gls{leo} satellite constellation operator, Sateliot and \gls{aws} have partnered to deliver cloud native \gls{5g} \gls{nbiot} service. This will enable Sateliot \glspl{leo} to act as \gls{5g} mobile transceivers and connect unmodified \gls{nbiot} devices to the \gls{5g} network globally \cite{Sateliot_5gNTN}. 
    \item Telstra and OneWeb: The two companies deliver \gls{leo} satellite-based cellular backhaul throughout Australia. Voice calls have been made through the OneWeb \gls{leo} constellation successfully, and the target of this collaboration to deliver 25 Gbit/s service to the remote mobile customers across Australia \cite{Telstra_5gNTN}.
    \item \gls{esa}, Telesat and Amarisoft: The alliance between the three parties has successfully materialised the world's first \gls{3gpp} (Release 17) \gls{ntn} link over \gls{leo}, between the ESTEC \gls{5g} laboratory and the Telesat \gls{leo}-3 satellite, taking a step forward towards delivering \gls{5g} over \gls{ntn}. The \gls{5g} bidirectional link between gNB and the user supports 3 bits/s/Hz over adaptive modulation selection from \gls{qpsk} to 64-\gls{qam} \cite{esa_5gNTN}.  
    \item Japan-Europe long-distance \gls{5g} network over satellites: This test network has incorporated Ku-band over \gls{geo} satellites to connect \gls{5g} networks between Japan and Europe, and they have evaluated 4K video, \gls{iot} data, and network control signals over this link. Progressing towards unified satellite-terrestrial networks, their findings provide evidence of successful \gls{5g} networks over satellite, with tolerable latency \cite{japan_5gNTN}. 
\end{enumerate}

\subsection{Cognitive Radios (CR)}

The \gls{itu}-R SM.2152 defined the \gls{cr} systems as a radio network setup which leverages technology to extract operational and environmental knowledge, and dynamically and autonomously adjusts its operational parameters and protocols based on the obtained information to achieve predefined objectives. 
This approach can be leveraged to enhance spectral efficiency, especially in congested bands. 
In the literature, \gls{cr} systems often follow the hierarchical classification of \gls{pu} and \gls{su}, aiming to grant \glspl{su} access to under-utilized spectrum without disrupting \gls{pu}s.
The initial phase of the cognitive process is to observe the spectrum and gather information about spectrum utilization. 
The second phase is spectrum analysis, where \gls{cr} evaluates the observed signal characteristics. 
Based on the analysis, the optimal unused frequency band is selected for transmission according to the feature requirements \cite{hassan2021survey}. 
\gls{cr} process then reconfigures the software parameters to change transceiver frequency, switch methodology, and energy transfer approach \cite{SS_survey1}.  
Literature has explored the concept of \gls{cr} along multiple axes such as regulation \cite{zhang2016cognitive, CR4}, applications \cite{CR6, CR7} and technology \cite{hoyhtya2012application, 9003405} with the leap of \gls{ai}/\gls{ml} and with the adoption of \gls{sdr} in the past decade~\cite{Wang2019}.

Due to the inherited complexity in \gls{cr} techniques, excessive standardization is paramount. 
The \gls{ieee} 802.22 standard, developed by the \gls{ieee} 802 LAN/MAN Standards Committee, is the first published standard on leveraging \gls{ieee} in licensed bands for \gls{wran} \cite{802_22}. 
It elaborates on utilizing geo-location and \gls{ss} to enable \gls{cr} within dynamic radio environments, by identifying and using unoccupied frequency channels without interfering \gls{pu} communications. 
Geographical locations are facilitated through a database of local licensed transmitters or \glspl{pu}, while the \gls{ss} techniques are leveraged to detect unused frequency channels for \glspl{su} to transmit. 
Further, \gls{cr} methodologies are extended to efficient use of unlicensed spectrum through \gls{ieee} 802.15 standard \cite{802_15}, which focuses on the coexistence of Wi-Fi and Bluetooth networks. 
In addition, the Dynamic Spectrum Alliance was formed to oversee the advancement of \gls{cr} standardization and deployments in 2014 \cite{ds_alliance}. 
It has contributed significantly towards \gls{tvws} utilization for broadband and developing \gls{cr} standards in collaboration with regulatory authorities. 

The following is a summary of materialized implementations of \gls{cr} approaches:
\begin{enumerate}
    \item \gls{tvws} in Rural Broadband: The widely adopted \gls{cr} technology for commercial deployments. According to the Dynamic Spectrum Alliance, more than 10 successful commercial or pilot projects have been completed worldwide, including the \gls{usa}, the \gls{uk}, Japan, Canada, and Ghana~\cite{ds_alliance}. 
    Another test project to utilize \gls{tvws} was successfully implemented in India, which aimed to provide broadband connectivity to rural areas at a low cost \cite{khaturia2018tv}. 
    Further, preliminary surveys have been done in the USA to regularize the use of \gls{tvws} \cite{nekovee2009survey}, thus providing broadband internet access in rural and underserved areas.     
    % \item Public Safety Communications: Cognitive radio has been deployed in dynamic spectrum access systems for public safety communications. In critical situations, such as natural disasters or emergencies, cognitive radios allow first responders to dynamically share spectrum resources, improving communication reliability and efficiency. This deployment ensures that emergency services have the necessary bandwidth for reliable communication when it is most needed.  
    \item \gls{darpa} XG Program: The Next Generation~(XG) program initiated by \gls{darpa} primarily focused on developing \gls{cr} technologies for military and public safety applications \cite{darpa}. 
    The program successfully demonstrated \gls{cr} capabilities under challenging radio conditions and developed an architectural framework and protocols for \gls{dsa}, enabling military communications systems to use unutilized spectrum more efficiently, avoiding interference with other users.    
    \item Nokia’s \gls{cr} Networks: Nokia has successfully carried out several \gls{cr} implementations for testing purposes \cite{mukherjee2015cognitive}. These tests focused on leveraging \gls{cr} techniques to enhance spectral usage in high-demand urban environments. 
    Furthermore, Nokia has developed a self-organizing network solution to dynamically change radio parameters to meet the user demand, thereby improving mobile network performance and user experience \cite{8368999}. 
    \item NASA SCaN: John H. Glenn Research Center in NASA, led by the \gls{scan} group, places a research platform on the \gls{iss}. This \gls{scan} testbed, equipped with three \glspl{sdr}, enables \gls{cr} research on orbital communication platforms \cite{ferreira2018multiobjective, reinhart2014space}
\end{enumerate}

\gls{cogsat} refers to the concept of adapting \gls{cr} capabilities to \gls{satcom} systems, enabling real-time \gls{ss}, dynamic access, and intelligent decision-making based on the environmental context and spectral occupancy. 
This enables satellites and heterogeneous space-terrestrial networks to identify underutilized spectrum bands and intelligently reuse them without causing harmful interference to primary/license users. 
Furthermore, enhanced \gls{cogsat} techniques tailored to encapsulate spatial, spectral, and temporal conditions allow adaptive link configuration, interference avoidance, and load balancing across beams and frequency bands. 
% Later discussions along the specific subtopics will provide a detailed overview of the \gls{cogsat} concepts explored with regard to \gls{dsm} in the context of application, technology, and regulatory aspects. 

\subsection{AI \& ML}

\gls{ml} is a subset of \gls{ai} that enables machines or systems to learn and improve from experience without being explicitly programmed for specific tasks. 
The taxonomy of \gls{ml} is complicated, considering the novel approaches branched out from the preliminary algorithms. 
Classifying \gls{ml} algorithms is a complex task \cite{mahesh2020machine}, however, in this work, we are detailing well-explored \gls{ml} methods and \gls{ml} approaches that carry the potential to significantly impact satellite spectrum management.  

\subsubsection{Supervised Learning}

Refers to a type of \gls{ml} methodology targeted at predicting or classifying outcomes using labeled data, where both input features and corresponding outputs are known. 
It is particularly effective for tasks requiring prediction or classification in dynamic environments such as satellite and terrestrial networks, for identifying signal patterns, resource allocation, or anomaly detection \cite{burkart2021survey}. 
% The general SL process involves training an \gls{ml} empowered with a deep neural network on a dataset to understand the hidden relationships between inputs (features) and outputs (targets). 
% The trained \gls{ml} model can then generalise its learning to new, unseen data for similar tasks, generating predicted outcomes. 
\gls{sl} plays an important role in enhancing decision-making and optimizing operations in dynamic and complex communication environments such as satellite networks \cite{fourati2021artificial}.
Key \gls{sl} algorithms explored in the literature are as below: 

\begin{itemize}
    \item Linear and Logistic Regression: 
    Used for predicting continuous outcomes (linear regression) or binary classifications (logistic regression). These are interpretable methods suited for simple relationships between features and targets~\cite{burkart2021survey}.
    
    \item Decision Trees: 
    Contrive a tree-like model to split data into branches/sections based on feature thresholds. Accountable and effective for both classification and regression~\cite{navada2011overview}.
    
    \item Random Forests: 
    An ensemble method that builds multiple decision trees and aggregates their predictions for robust and less over-fitted results~\cite{liu2012new}.
    
    \item \gls{svm}: 
    Separates classes by identifying the hyperplane with the maximum margin between different categories. Effective for high-dimensional data but less interpretable~\cite{tang2013deep}.
    
    \item Regularization Techniques: 
    Introduces penalties to regression models to reduce over-fitting and handle multicellularity among features. LASSO, Ridge, and Elastic Net are a few regularization techniques found in literature \cite{ nusrat2018comparison}.
    
    \item Ensemble Learning: 
    Also referred to as Super Learning, combines multiple algorithms to optimize prediction performance by leveraging the strengths of each individual method \cite{dong2020survey}.
    
    \item \gls{knn}: 
    Classifies a sample based on the majority class of its nearest neighbors in feature space, suitable for simpler, smaller datasets \cite{ML3}.
\end{itemize}

\subsubsection{Unsupervised Learning}

Refers to a \gls{ml} approach used to identify patterns and structures in unlabeled data \cite{sindhu2020survey}, which is particularly beneficial for environments such as satellite networks where vast amounts of data are generated without pre-classification. 
% Using techniques such as clustering, association rule learning, anomaly detection, and autoencoders, unsupervised learning can uncover hidden insights within satellite communication systems. 
An overview of key unsupervised learning algorithms is as follows:

\begin{itemize}
    \item Clustering: 
    Group data points based on similarity. A widely used algorithm that falls under this is K-Means. This approach is often used to identify patterns in unlabeled datasets \cite{ camino2021milp}.

    \item Association Rule Learning:
    As the name implies discovers relationships, associations, and patterns between variables. Apriori algorithm is a well-explored method, which can be categorized under this method \cite{al2014improved}. 

    \item Anomaly Detection:
    Identifies irregular patterns in data, useful for detecting signal interference or system faults~\cite{gunn2018anomaly}.

    \item Autoencoders:
    Neural networks specifically designed for dimensionality reduction and feature extraction. This approach has unique advantages in efficiently processing large datasets \cite{saifaldawlaconvolutional}.

    \item \gls{pca}: 
    Reduces data dimensionality while retaining key information. Further enabling faster and more accurate processing for resource optimization and fault analysis \cite{vazquez2021machine}.
\end{itemize}

In the context of satellite communication, clustering algorithms can group satellite transmission data or identify spectral usage patterns. Anomaly detection approaches can monitor signal behavior irregularities or identify jamming attempts. 
Adapting these \gls{usl} methods enables autonomous decision-making, allowing systems to adapt instantly to dynamic environments without human intervention \cite{saifaldawlaconvolutional, li2021scalable}.
The flexibility and scalability inherited by unsupervised learning techniques make the methods well-suited for diverse applications, such as resource management, fault detection, and data optimization across various domains \cite{vazquez2021machine, sindhu2020survey, mahesh2020machine}.
Considering the vast array of data complex network environments generate, advanced techniques like dimensionality reduction with \gls{pca} can enhance data processing efficiency by ensuring that models focus on the most relevant features of the data.
% Therefore, integration of these unsupervised learning methods enables systems to improve operational efficiency, enhance resilience, and optimize performance in complex and dynamic environments such as satellite communication networks.

\subsubsection{Reinforcement Learning}
Refers to a feedback-based methodology of \gls{ml}, in which a learning agent takes actions in an environment based on the rewards offered for the action.
For each preferred action which the agent takes in a particular state, it gets positive feedback, and bad actions gets penalized. 
The agent's transition from one state to another depends on the previous state, the action taken, and the next state. 
There is no labeled training data in \gls{rl}, therefore, the agent is bound to learn from its experience. 
Hence, the agent reacts with the environment and explores by itself while trying to maximize cumulative rewards \cite{kaur2022comprehensive}. 
When complete information about the system is available, the dynamic programming approach can be used to determine the optimal policy. 
The formal model of \gls{ml} is the \gls{mdp}, however, when complete system information is not available other than the sequence of past states, actions, and rewards, the Monte-Carlo method can be applied to get the optimal policy. 
Temporal difference learning takes a different approach to the above by not forming a system model \cite{graesser2019foundations}. 

\gls{drl} adds \gls{rl} and \gls{dl} techniques together, enabling agents to make decisions in the state space using unstructured input data. 
Due to the capability of \gls{drl} to take large inputs, and the multiple neural network architecture makes \gls{drl} an ideal methodology to exploit real-world problems and achieve resolutions beyond human capabilities \cite{graesser2019foundations}.
\gls{madrl}, focuses on the behavior of multiple learning agents in a shared environment. 
In \gls{madrl}, each agent is responsible for its own actions and rewards, often pursuing different and sometimes conflicting objectives, leading to complex group dynamics. 
Compared to conventional \gls{drl} methods, \gls{madrl} is compatible with information sharing between agents. This helps in accelerating the learning of similar tasks and enables the achievement of better overall performance. 
In addition, when an agent or several agents fail, the remaining agents can take over their tasks, demonstrating the inherent robustness of this approach.

\subsubsection{Distributed Learning}

Distributed learning is a paradigm of \gls{ml}, where the workload is spread and shared across multiple nodes, which enables the processing of large data sets under time and computational resource constraints \cite{verbraeken2020survey}. 
Parallelism is the key concept behind distributed learning, which facilitates data splitting across multiple nodes where each node processes a portion of the data. 
In addition, model parallelism enables dividing the \gls{ml} model across different nodes where each node computes parts of the model. 
Under realistic settings, both terrestrial and non-terrestrial radio networks are wide-area networks with distributed computational resources. 
Processing large datasets generated through these networks demands higher computational resources, and the end nodes are inherently computationally constrained, thus making distributed learning an ideal approach for  \gls{satcom} networks. 
Further, utilizing the limited communication resources to transmit sensory data to build the \gls{ml} model affects the total system efficiency. 
\gls{fl} is a well-established distributed learning approach that can counter the disadvantages of centralized learning approaches. 
In \gls{fl}, data remains decentralized and only the \gls{ml} model updates are shared with a central server, thus reducing the communication overhead while enhancing privacy and security \cite{ML1, razmi2024board}.

\subsubsection{Generative AI and Large Language Models}

Generative \gls{ai} and \glspl{llm} represent groundbreaking advancements in \gls{ai}, enabling machines (essentially large deep neural networks) to understand, process, and generate human-like content with remarkable accuracy and coherence \cite{yao2024survey}. 
\glspl{llm}, such as OpenAI’s \gls{gpt} series \cite{roumeliotis2023chatgpt} and Meta's llama \cite{touvron2023llama} are a class of \gls{dl} models built on the transformer architecture.
These models consist of a large neural network framework that efficiently encapsulates long-range dependencies in sequential data through mechanisms like self-attention. 
% These LLMs are pre-trained on a broad corpus of text data from diverse sources, such as books, articles, websites, and code repositories, allowing them to learn intricate patterns of grammar, semantics, and even reasoning. 
On the other hand, Generative \gls{ai}, a broader category encompassing \glspl{llm}, focuses on creating new content imitating the style and data structure it was trained on. 
Generative \gls{ai} models extend to producing images, music, video, and even synthetic datasets, utilizing models such as \glspl{gan} \cite{goodfellow2020generative} and \glspl{vae} \cite{kingma2019introduction}, expanding beyond text generation.
% Natural Language Processing~(NLP) is highly influenced by the revolution of LLMs and generative AI, creative content generation, chatbots, and virtual assistants.
% These technologies enable computers to perform tasks from answering questions, summarizing text, generating conversational responses, expanding even to writing code and creating fictional stories extracting human generative qualities.
% Zero-shot or few-shot learning, which refers to Generative AI's ability to generalize across tasks with no additional training makes them incredibly versatile. 
% However, in order to achieve this level of superiority, deep neural networks with billions of parameters need to go through an extensive training process, which requires significant computational resources.
Moreover, as these \glspl{llm} continue to evolve, their impressive performance is adapted to robotics in generating control commands \cite{driess2023palm}, in biology for predicting protein structures \cite{lin2023evolutionary}, and in networking \cite{wu2024netllm}.
Such cross-domain adaptations have initiated a promising avenue for \glspl{llm}, which has significant potential to enhance operations in unexplored areas such as \gls{satcom}. 
% The combination of generative AI and LLMs represents not just a technological shift but a profound transformation in how humans interact with machines, unlocking new possibilities for creativity, problem-solving, and automation.

\subsection{Software-Defined Radio and Networking}

\subsubsection{Software-Defined Radio (SDR)}
\gls{sdr} is a versatile radio communication approach that uses software to define and control radio frequency functionalities such as modulation, demodulation, signal processing, transmission power, and frequency selection \cite{jondral2005software}. 
Unlike traditional hardware-based radios with fixed parameters, \gls{sdr} relies on reprogrammable components, such as \glspl{fpga} and \glspl{gpp}, to adapt its operation dynamically according to the environment of operation. 
This flexibility in \gls{sdr} makes it an essential enabler of \gls{cr}, empowering radios to modify their parameters in real-time based on environmental conditions, user demands, and regulatory requirements \cite{jondral2005software}. 
\gls{sdr}'s ability to reconfigure its transmission characteristics paves the way for advanced spectrum management techniques, thus making it foundational for enabling adaptive and intelligent capabilities that are vital for \gls{cr} systems.

In the context of \gls{cogsat} networks, \gls{sdr} serves as the hardware platform that supports \gls{dsa} and real-time environmental sensing, enabling environmental dynamic-based decision making and automated changes. 
For instance, \glspl{cr} requires the ability to detect and exploit spectrum holes (discussed in Section \ref{sec: dsm for cogsat}) to dynamically adjust transmission parameters and switch between frequency bands under pre-programmed guidelines. 
\gls{sdr}'s programmability and reconfigurability enable these functions, allowing seamless transitions across access communication protocols and frequencies. 
For example, a \gls{cr} with an \gls{sdr} platform can enable flexible transmission from a Wi-Fi band to a cellular band when spectrum congestion and other anomalies are detected, ensuring uninterrupted communication \cite{gudipati2013softran}. 
Furthermore, \gls{sdr} integration enables realizing advanced signal processing algorithms in \gls{ss} and modulation recognition, which are crucial for identifying opportunities and mitigating interference in shared spectrum environments.

\subsubsection{Software-Defined Networking (SDN)}

\gls{sdn} introduces a paradigm shift by decoupling the control and data planes, thereby centralizing decision making \cite{jiang2023software}.
Integrating these approaches in \gls{satcom} systems play a pivotal role in enabling intelligent and \gls{dsm} \cite{li2016using,qi2022sdn}.
% Separation of control and data planes enables compact and lightweight satellites, such as \glspl{leo}, to focus on lightweight packet forwarding, while the \gls{sdn} controllers manage complex tasks such as spectrum allocation, interference mitigation, routing optimization, and dynamic reconfigurations.
% \gls{sdn} controllers can be distributed throughout the constellation, which are typically deployed in ground stations or selected centralized satellites \cite{jiang2023software}. 
\gls{sdn} brings indispensable qualities to \gls{satcom} environments in managing current mega satellite constellations, which operate under a highly dynamic and time-varying network topologies. 
By maintaining a global view of the network, including channel allocations, \gls{isl} conditions, and traffic congestion, \gls{sdn} enables adaptive and efficient spectrum allocation policies, fulfilling key intelligent spectrum management objectives.
Moreover, \gls{sdn} is a primary enabler of seamless integration of satellite and terrestrial networks into a unified, programmable infrastructure, enabling cross-domain policy enforcement, such as coordinated spectrum reuse and terrestrial to satellite network handovers \cite{LP5}. 
Its programmable interfaces support the integration of \gls{ai}/\gls{ml}-driven spectrum management, getting the advantage of real-time telemetry for context-aware decision making. 
% This enhances fine-grained network slicing modules dedicated to dynamic frequency band assignment across multi-orbital satellite network layers based on service demands and orbital configurations, such as \gls{geo}-\gls{leo} coexisting networks. 
Furthermore, the multi-domain and hierarchical controller architecture of \gls{sdn} ensures scalability and supports agile reconfiguration in response to satellite mobility and link disruptions \cite{jiang2023software}. 
% Therefore, \gls{sdn} lays the technological foundation for intelligent, adaptive, and efficient spectrum management strategies necessary to meet the diverse \gls{qos} requirements in the ever-growing satellite networks.

\subsection{Network Function Virtualization (NFV)}

% \gls{nfv} is a transformative enabler for intelligent spectrum and resource management in \gls{satcom} networks. 
Traditional satellite infrastructures are characterized by monolithic, vendor-specific hardware, which constrains flexibility, increases operational costs, and limits the integration of new services and protocols. 
\gls{nfv} addresses these limitations by separating network functions, such as routers, firewalls, performance-enhancing proxies, and load balancers, from proprietary hardware, allowing them to run as \glspl{vnf} on commodity computing platforms \cite{jia2021vnf}. 
% This virtualisation encourages dynamic provisioning, agile deployment, and scaling of services across multiple domains, enabling real-time response.
These precedents can be leveraged to expand mega satellite constellations catering to dynamic demands \cite{maity2024traffic}. 
Especially in satellite-terrestrial hybrid networks, \gls{nfv} enhances operational efficiency by allowing service providers to dynamically instantiate and orchestrate \glspl{vnf} that can adaptively manage spectrum usage based on traffic, interference temperatures, and \gls{qos} metrics.  
Through such virtualization, intelligent spectrum management becomes feasible via on-demand resource allocation, fast reconfiguration, and improved interoperability with terrestrial networks.
% Furthermore, \gls{nfv} empowers satellite operators to adopt more versatile business models such as Satellite Virtual Network Operators (SVNOs), where infrastructure resources are logically partitioned and leased to third-party service providers. 
% In this context, \gls{nfv} allows SVNOs to deploy dedicated VNFs, and create satellite network slices without needing full ownership of the physical satellite infrastructure \cite{jia2021vnf}. 
% This not only optimizes the utilization of radio spectrum and network resources but also supports differentiated service offerings tailored to specific applications and customer segments. 
\gls{nfv}, in cooperation with \gls{sdn}, facilitates distributed control, multi-tenancy, and service chaining across satellite and terrestrial segments, thus enabling end-to-end resource management intelligence. 
Such architectures support the introduction of use cases such as on-demand bandwidth allocation and edge processing, which are key to adaptive and efficient spectrum use in dynamic operational environments \cite{LP9}. 
% Ultimately, \gls{nfv} underpins a shift from static, rigid satellite communication systems to flexible, service-oriented infrastructures, paving the way for realizing intelligent and autonomous spectrum management.

\subsection{Edge Computing}

This methodology pushes computation and decision-making closer to where data is generated.
\gls{mec}, an extension of edge computing, enables data processing with reduced hops in communication networks. 
In \gls{satcom} networks, edge computing clusters can be located in satellites, distributed earth stations, or they can take a hybrid form, scattering and processing data between satellite-earth station hybrid systems \cite{zhang2019satellite}. 
Satellite networks relying on centralized cloud-based processing introduce significant latency and bandwidth overhead, especially for delay-sensitive applications, due to the inherited signal traveling distances. 
In contrast, through the satellite-terrestrial integrated edge computing networks and \gls{leo}-based edge computing paradigms, edge computing reduces the number of data traveling hops, thus minimizing the latency \cite{xie2020satellite}. 
These advantages can be leveraged for spectrum analytics, local policy enforcement, and \gls{dsm} directly as they demand real-time configuration changes, considering dynamics in \gls{satcom} networks. 
Therefore, edge computing is an important facilitator in reducing processing latency and back-hauling traffic, while enhancing spectrum responsiveness and autonomy in \gls{satcom} systems. 

% Furthermore, integration of edge computing enables localised caching and context-aware decision-making, which are crucial for supporting dynamic and intelligent spectrum access in scenarios with intermittent connectivity or remote user distribution. 
% In addition, edge computing in \gls{cogsat} environments empowers predictive spectrum planning by leveraging onboard processing capabilities through processing \gls{ss} data with interference management. 
% Moreover, MEC-enabled satellite-terrestrial architectures support multi-layer and heterogeneous clusters, including terrestrial MEC, satellite-based edge nodes, and aerial platforms that collaboratively handle computation-intensive tasks \cite{xie2020satellite}. 
% These platforms can offload and schedule tasks across nodes based on real-time link and resource status, ensuring low-latency access and resilient operations. 

\subsection{Blockchain for privacy and security}

Blockchain is emerging as a transformative enabler of intelligent spectrum management, as it addresses challenges related to security, decentralization, transparency, and automation of spectrum transactions~\cite{liang2020dynamic}. 
The inherent security within blockchain architecture offers a decentralized and tamper-resistant database, such as \gls{rem}, for spectrum usage records, enabling transparent and secure sharing of spectrum access among stakeholders. 
In addition, blockchain facilitates \gls{dsm} policy enforcement through smart contracts without relying on centralized authorities \cite{chen2022blockchain}. 
This facilitates the creation of a self-organized spectrum market that supports real-time trading and leasing of unused spectrum, optimizing utilization across satellite networks. 
Moreover, blockchain secures \gls{ss}, spectrum auctions, and dynamic access processes, enhancing trust among participants in a decentralized \gls{cogsat} ecosystem~\cite{perera2024survey}.

\section{Regulations and standardizations in Satellite Spectrum Management}
\label{sec: regulations}

This section discusses the existing regulatory and standardization bodies on \gls{satcom} networks in the context of spectrum management and satellite network integrations, with their contributions towards the advancement of the \gls{satcom} industry.

\subsection{IEEE}
As a leading standardization body,  the \gls{ieee} has played a significant role in advancing \gls{satcom} networks by developing technical standards that promote efficient spectrum utilization and interoperability.
The frequency bands allocated to \gls{satcom} generally fall within the 1–40 GHz range, although application-specific satellites may operate outside this spectrum.
The \gls{ieee} categorized them into seven frequency bands with their own characteristics and properties, making them suitable for specific satellite operations. 
% Table \ref{satellite frequencies} provides an overview of the \gls{satcom} bands and their applications \cite{gagliardi2012satellite, gps}. 
These \gls{satcom} bands and their applications are discussed extensively in \cite{gagliardi2012satellite}. 
The lower part of the spectrum has higher propagation qualities and, therefore, is utilized in applications with extensive coverage and low throughput requirements. 
Higher frequencies can facilitate more bandwidth and require higher transmission power to compensate for signal degradation. 
% \gls{leo} satellites are generally deployed for use cases with large capacity requirements, focusing on broadband applications that can facilitate different applications:
% \begin{itemize}
%      \item Mobile backhauling where terrestrial networks are expensive to deploy.
%      \item Broadband for mobile use cases that require terrain coverage \textit{e.g.}, naval transport vessels, and aeroplanes.
%      \item Fixed broadband for remote and underserved areas. 
%      \item Dedicated global enterprise connectivity. 
% \end{itemize}
Therefore, most \gls{leo} satellites utilize Ku (12–18 GHz) and Ka (26.5–40 GHz) frequency bands to facilitate high-throughput connections, as they offer higher bandwidth and data rates. 
Additionally, the shorter wavelengths allow for smaller, lighter antennas and more efficient beam forming, which are advantageous for both satellite payload design and user terminals.

\begin{comment}
\begin{table}[ht]
\caption{{Satellite Frequency bands and their applications.}}
\centering
\resizebox{0.4\textwidth}{!}{
\begin{tabular}{|p{0.6cm}|p{1.9cm}|p{5cm}|}
\hline
Band & Frequency Range (GHz) & Applications \\
\hline
L & 1-2 & Narrowband communication.  \\
S & 2-4 & Weather and ship radars, SAT based digital radio broadcasting.   \\
C & 4-8 & SAT TV broadcasting.     \\
X & 8-12 & Military-based SAT applications.   \\
Ku & 12-18 & VSAT~(Very Small Aperture Terminal) system communication, and broadcast SAT communications. \\
K & 18–27 & Only used in short-range applications.\\
Ka & 26-40 & High throughput SAT communications.  \\
V & 40-75 & Inter-SAT and high throughput SAT communications. \\
W & 75-110 & High throughput SAT communications. \\
\hline
\end{tabular}
}
\label{satellite frequencies}
\vspace{-2mm}
\end{table}
\end{comment}

\gls{ieee} has played a crucial role in integrating \gls{satcom} with \gls{5g} and beyond next-generation wireless networks \cite{fu2023satellite}. 
Established \gls{ieee} \gls{5g} and \gls{6g} working groups focus on defining the coexistence of terrestrial and satellite networks, particularly in shared and mmWave spectrum bands. 
\gls{ieee}’s drive on research for beam-forming, \gls{mimo}, and \gls{ai}-driven spectrum management, enhancing the spectral efficiency of satellite-based \gls{ntn}. 
Furthermore, \gls{ieee}'s collaboration with regulatory and standardization bodies such as \gls{itu} and \gls{3gpp} ensures that its efforts align with global telecommunications frameworks for seamless network integration.
One of the most notable contributions towards intelligent spectrum management that can be adapted to \gls{satcom} platforms is the \gls{ieee} 1900 series through Dynamic Spectrum Access Networks~(DySPAN) Standards Coordinating Committee 41~(SCC41), which focuses on \gls{dsa} and \gls{cr} systems critical for spectrum efficiency in satellite and terrestrial networks. 
It further defines key concepts in \gls{cr}, policy-based radio, adaptive radio, and \gls{sdr}, ensuring efficient spectrum utilization and interoperability, with technical guidelines for analyzing coexistence and mitigating interference between radio systems operating in overlapping or adjacent frequency bands.
By adopting real-time spectrum monitoring, interference prediction, and adaptive mitigation strategies, \gls{ieee} 1900 further enhances spectrum-sharing efficiency for applications such as \gls{cogsat} communications and terrestrial-satellite hybrid networks.

Additionally, \gls{ieee} 802.16 for wireless broadband access has been adapted for \gls{satcom} applications \cite{IEEE80216}, enhancing internet accessibility in remote areas. 
Protocols such as \gls{ieee} 802.22 for \gls{wran} \cite{IEEE80222} have also been instrumental in utilizing TV white spaces, which can be leveraged for satellite-terrestrial hybrid networks. 
Furthermore, \gls{ieee} has also made significant contributions to the global rise of \gls{iot} over satellite through standardizing low-power communication protocols \cite{centenaro2021survey}.
IEEE’s 802.11 \gls{ai}/\gls{ml} topic interest group focuses on the application of \gls{ai} in wireless networks, as such developments carry the potential to optimize connectivity through intelligent spectrum allocation and interference mitigation. 
% IEEE also drives collaborations with other standards bodies such as ITU and 3GPPP, advancing \gls{satcom} standards, NTN integration, and IoT connectivity, playing a pivotal role in the evolution of satellite networks. 

\subsection{ITU}
The \gls{itu} is the United Nations specialized agency for digital technologies. \gls{itu} coordinates global spectrum and satellite orbit usage via the radio regulations treaty and develops the technical standards that ensure networks and technologies connect seamlessly \cite{ITU_regulations, sharma2013satellite}.
As discussed above, satellite transmission frequency offers unique characteristic advantages that can be leveraged for different use cases. 
To this end, \gls{itu} has significantly contributed to characterizing the environmental factors, channel fade models, and dynamics, improving \gls{satcom} service efficiency.
Some of the key points \gls{itu} that have been addressed in satellite link modeling are as below:
\begin{itemize}
    \item Absorption, scattering, and depolarization by water, ice drops, clouds, and other hydrometeors in the atmosphere. 
    \item Signal loss due to the refraction in the atmosphere. 
    \item Antenna gain decreases due to the phase decorrelation caused by irregularities in the refractive index. 
    \item Slow fading caused by beam bending, and rapid fading due to refractive index variations. 
    \item Bandwidth limitations caused by multipath scattering. 
    \item Varying elevation angle of \gls{leo} satellites. 
    \item Faraday rotation and ionospheric scintillation.    
\end{itemize}
In addition to discussing the above factors extensively in \cite{ITU2}, \cite{ITU1} studies tropospheric and ionospheric effects, shadowing in satellite-to-land channels with measurements up to 20 GHz. 
Furthermore, the \gls{itu} studies present multi-path channel models for clear \gls{los} conditions, a statistical model for mixed propagation conditions, and a physical statistical wide band model for mixed propagation conditions.

Through its radio regulations for international frequency management, the \gls{itu} establishes global policies that prevent harmful interference between satellite services by assigning specific frequency bands to various satellite applications/ operators \cite{ITU_regulations}.
To manage frequency assignment and satellite characteristics, the \gls{itu} categorizes the existing satellite services into three divisions as \gls{fss}, \gls{mss}, and \gls{bss}. 
Furthermore, the \gls{itu} maintained \gls{mifr}, ensuring the global recognition of frequency assignments, facilitating structured and interference-free spectrum management across borders \cite{mifr}.
In addition, the \gls{wrc}, hosted by the \gls{itu} every four years, acts as the primary forum for updating regulations to address emerging technological needs.
In the last \gls{wrc} held in 2023, coexistence between \gls{ngso} and \gls{geo} satellites with established power limits for \gls{ngso} satellites, exploring new frequency bands for mobile satellite services, and equitable access to spectrum for developing countries were among the key points discussed \cite{ITU_wrc23}.

The \gls{itu} constitution acknowledges that radio frequencies and satellite orbits are limited natural resources, which necessitates rational, efficient, and economic usage to benefit both developed and developing nations.
The coordination and regulatory framework \gls{itu} has established for satellite networks further strengthens the fair usage of \gls{satcom} spectrum, balancing the efficient utilization of spectrum resources with equitable access for all countries.
In order to materialize this fair usage policy, \gls{itu} leverages a ``first-come, first-served" coordination procedure, ensuring the orbital and spectral resources are allocated based on actual needs, thus improving spectrum utilization. 
Additionally, \gls{itu} has introduced planning mechanisms that reserve frequency allocations for future use, particularly safeguarding access for nations that do not have a presence in orbit and \gls{satcom}, thus ensuring that the spectrum is equitably distributed.
To this end, \gls{itu}'s radio regulations framework establishes clear rules for frequency coordination, advance publication, and notification of \gls{satcom} networks, ensuring fair access while encouraging further advancements.

The \gls{itu} has progressively adapted its regulatory framework to support \gls{ai}-driven spectrum management as it believes \gls{ai} approaches to facilitate resource management in \gls{satcom}, such as coverage adjustments, capacity, and spectrum allocation. 
With \gls{itu} reports such as ITU-R S.2357 and ITU-R S.2361, the organization contributes to providing guidelines for naval \gls{ai} aspects in \gls{fss} communications with mobile platforms and broadband access, respectively. 
Beyond allocation and coordination, the \gls{itu} has been instrumental in defining regulations for emerging satellite technologies, such as \gls{ngso} satellites, \gls{5g} and beyond \gls{ntn}, and satellite-based \gls{iot}, in collaboration with other standardizing bodies, such as the \gls{3gpp}. 
The \gls{itu}’s role in space sustainability is evident in its efforts to prevent spectrum congestion, reduce signal interference, and implement space debris mitigation policies. 
By continuously evolving its regulations, \gls{itu} remains at the forefront of global satellite spectrum governance, ensuring its sustainability and accessibility for enhanced \gls{satcom} networks. 

% Based on the ITU regulations non-\gls{geo} satellite systems shall not cause unacceptable interference to \gls{geo} satellite networks \cite{christensen2012itu}. 

\subsection{3GPP}

\gls{3gpp} is the prominent standardization body for terrestrial communication; however, with the recent advancements in the \gls{satcom} sector, it also has significant contributions towards satellite and \glspl{ntn}. 
Consequently, \gls{3gpp} has been instrumental in standardizing frequencies and related technologies for terrestrial networks, thereby enabling the seamless integration of \gls{satcom} into terrestrial mobile communication systems.
S, L, and Ka are among the key spectrum bands the \gls{3gpp} has identified for \gls{ntn} communication, considering user requirements and spectral characteristics.
These standardized frequency allocations align with \gls{itu} regulations, ensuring global spectrum harmonization and preventing interference. 
\gls{3gpp} Release 18 is a key reference guideline to overcome hurdles in standardizing terrestrial and non-terrestrial networks, creating a common platform for both architectures.
With Release 18, the \gls{3gpp} has encapsulated interference mitigation techniques like power control, beam-forming, and \gls{dsa} into its \gls{ntn} specifications, aiming to manage interference effectively within satellite-based communication systems, particularly in the context of \gls{5g} and beyond networks \cite{3GPP_r18, 3GPP}. 
Furthermore, enhancements in waveforms, timing synchronization, and power management are introduced in \gls{3gpp} Release 18 to address challenges such as Doppler shift and latency, making \gls{satcom} networks more compatible with terrestrial network infrastructure.

\gls{3gpp} Release 18 also makes significant contributions towards the \gls{ntn}-terrestrial network integration, marking the first formal standardization support for mobile cellular network extension through \gls{ntn} \cite{3GPP_r18}. 
In reference to integrations with \gls{geo} and \gls{leo} constellations, \gls{3gpp} identified several implementation scenarios based on the payload, which are extensively discussed in ~\cite{vanelli20205g}. 
Primarily, these \gls{3gpp} architecture options can be categorized as transparent and regenerative. 
Transparent payloads can support multiple functions and user equipment without physical modification of the data packets, on the other hand, regenerative payloads are altered to match the transport network. 
Thus, focus has shifted to transparent payload options since they can harness the advantage of \glspl{isl}. 
In addition, user equipment can be connected to a relay node, rather than directly connected to the satellite network, which can be a popular adaptation considering power and radio connectivity limitations.

\begin{comment}
\begin{table}[]
\caption{{3GPP reference scenarios of Satellite-Terrestrial Network Integrations.}}
\begin{center}
\resizebox{\columnwidth}{!}
{
\begin{tabular}{|p{5cm}|p{2cm}|p{2cm}|}
\hline
Satellite System & Regenerative Payload & Transparent Payload \\
\hline
\gls{geo} & A & B \\ \hline
\gls{leo} steerable beams & C1 & D1 \\ \hline
\gls{leo} fixed beams & C2 & D2 \\ \hline
\end{tabular}
}
\end{center}
\label{3gpp}
\end{table}
\end{comment}

\begin{figure}[!t]
    \centering
    \includegraphics[width=0.95\columnwidth]{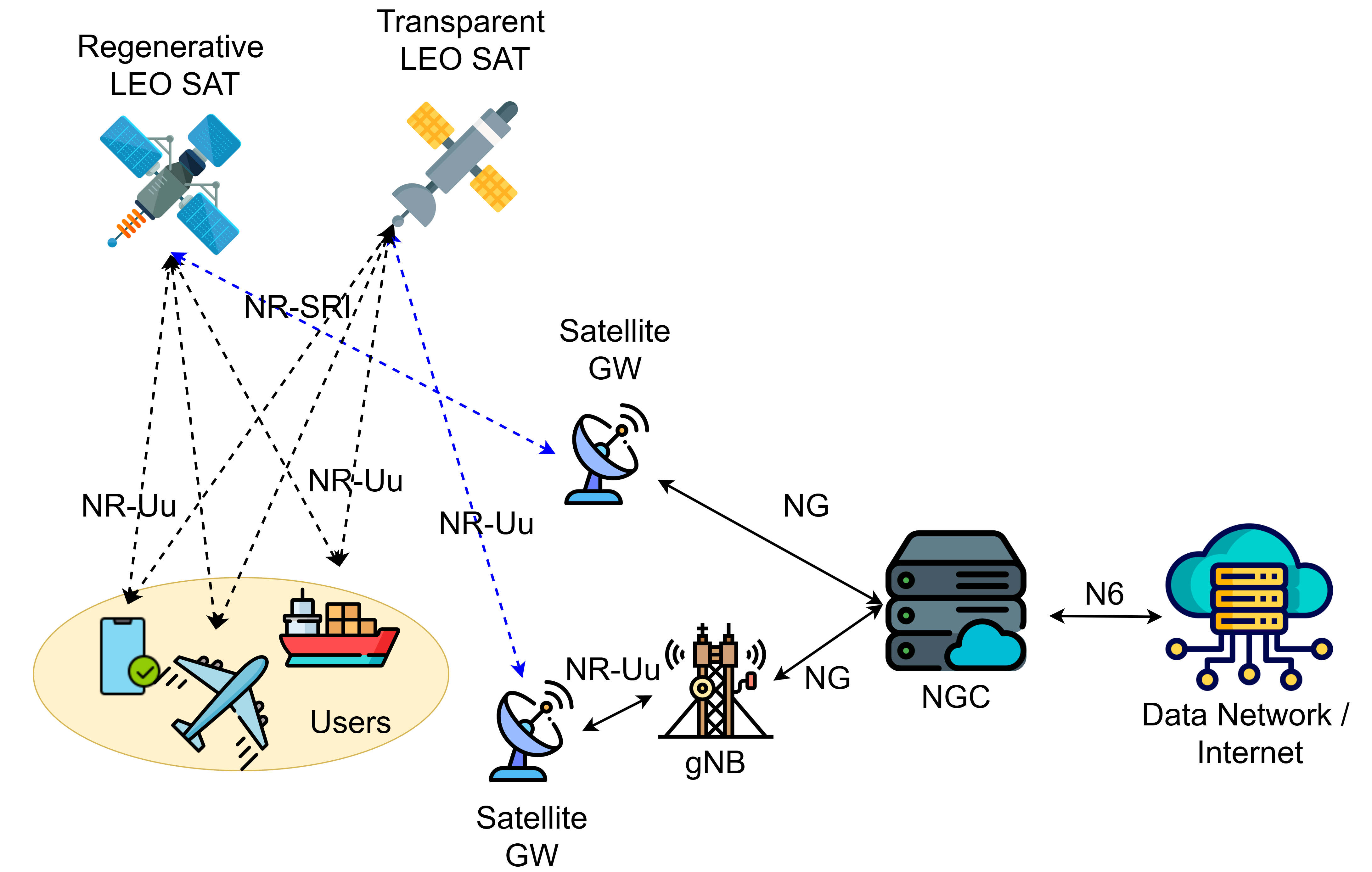}
    \caption{3GPP Satellite-Terrestrial integration architecture.}
    \label{sat_ter}
    \vspace{-5mm}
\end{figure}

Fig. \ref{sat_ter} illustrates the proposed \gls{3gpp} network architecture of satellite-terrestrial network integration for the direct access scenario. 
In the case of regenerative traffic, satellites should have the gNB functionality built onboard; thus, the gNB establishes an air interface with the gateway to route traffic to the Next Generation Core network~(NGC). 
Satellites can handle transparent traffic without inbuilt gNB functionality, and the received traffic is transported through a gateway to the terrestrial network. In this case, the next node can be a gNB. 
Regardless of the payload type, user equipment should access the satellites using an NR-Uu air interface to establish a proper data link layer connection. 

\gls{3gpp} contributions are extended in \gls{iot} and \gls{mtc} over satellite networks, enabling \gls{nbiot} and LTE-M to function in \gls{ntn} environments.
\gls{3gpp} has ensured spectrum efficient deployments of satellite-based \gls{iot} solutions by defining technical specifications for low-power, wide-area \gls{iot} communications over satellites. 
This enables \gls{iot} applications to extend beyond terrestrial coverage to support applications in maritime, remote agriculture, and disaster response.
These industries benefit from the real-time monitoring and connectivity \gls{satcom} networks provide through \gls{nbiot} 
Furthermore, with the adaptive modulation and coding techniques, \gls{3gpp}’s standards dynamically adjust to spectrum conditions, optimizing data transmission efficiency.
Through these policy developments and standardizations, \gls{3gpp} has significantly contributed to the integration of satellite services into mobile networks, expanding connectivity to remote and underserved areas while maintaining efficient spectrum utilization, thus paving the way towards global connectivity \cite{3GPP_NTN}.

\subsection{ETSI \& Europe}

The \gls{etsi} is another primary standardization body that plays a pivotal role in the context of \gls{satcom} systems, services, and applications, including fixed, mobile, broadcasting and satellite navigation systems.  
Its contributions towards satellite-terrestrial integration architecture can be considered as a key enabler in industrial satellite-\gls{5g} collaborative networks. 
Through its Satellite Earth Stations and Systems~(SES) technical committee, \gls{etsi} develops standards for \gls{satcom} networks, as well as satellite navigation and earth station equipment.
\gls{etsi} has further contributed to standardizing \gls{bsm} communications, enabling efficient IP-based satellite access networks.
The modular \gls{bsm} architecture, as outlined in \gls{etsi} TS 102 292 and TR 101 984, allows the integration of satellite-dependent transmission technologies with common satellite-independent IP networking functions, such as \gls{qos}, multicasting, and security.
\gls{etsi} contributions in this context enhance the efficiency of satellite communications through optimizing IP interoperability and compatibility between satellite and terrestrial networks, thus supporting next-generation broadband and \gls{ntn} advancements.

\gls{etsi} TR 103 611 recommends the fundamental standards in the integration of satellite and terrestrial networks within the \gls{5g} ecosystem, enabling seamless connectivity. 
Within it, \gls{etsi} defines architectural models and access scenarios, such as direct, indirect, and mixed \gls{3gpp} access of \gls{ntn} integration with terrestrial \gls{5g} systems.
The report further ensures interoperability between the two networks of interest.
By integrating \gls{haps} and satellites into \gls{5g}, \gls{etsi} promotes flexible, multi-layered network architectures capable of handling diverse communication needs, from \gls{embb} to \gls{mmtc} and \gls{urllc}.
In addition, \gls{etsi} TR 103 124 focuses on defining satellite-terrestrial network integrations. 
In that recommendation, \gls{etsi} emphasizes the rationale behind integrated networks and the value additions it can bring to the identified combined satellite-terrestrial network integrations. 
These integration standards open doors for co-existence of \gls{ntn} and terrestrial networks, thus highlighting the importance of efficient resource utilization in terms of spectral and computational assets. 
\gls{dsm} approaches allowing adaptive spectrum sharing between satellite and terrestrial systems while minimizing interference can play a pivotal role in these co-existing network setups.

Apart from the \gls{etsi}, Europe has several other authorized bodies that regulate and standardize \gls{satcom} operations. 
The \gls{cept} is one such organization, and they work closely with the \gls{itu} and \gls{etsi} in developing harmonized spectrum policies for European countries. 
\gls{cept}'s electronic communication committee defines policies for satellite spectrum management, including frequency coordination and interference mitigation for the European region. 
The \gls{ebu} is another regulatory body that works within the European region, and it sets standards and policies related to the use of satellite networks for data transmission within its jurisdiction. \gls{ebu} is responsible for standardizing satellite-based broadcasting services, including \gls{dth} TV, satellite radio, and multimedia distributions. In addition, \gls{ebu} also works on satellite frequency coordination for broadcasting applications across Europe.

\subsection{National and Regional Authorities}

In addition to the global framework established by the \gls{itu}, national and regional regulatory authorities impose guidelines on satellite spectrum management. 
They are often affiliated with national security interests and political agendas, considering the surveillance and monitoring capabilities satellites possess \cite{ sharma2013satellite}. 
Each country, through its national telecommunications authority, such as the \gls{fcc} in the United States, the \gls{acma}, or Ofcom in the United Kingdom, manages and licenses spectrum use within its jurisdiction.
These authorities ensure that satellite operators comply with national laws and international obligations, often enforcing licensing conditions and monitoring for interference.
On a regional level, organizations such as the \gls{cept}, the \gls{citel}, and \gls{apt} facilitate cooperation among neighboring countries and study related legal issues, coordinating spectrum use to minimize cross-border interference.
Furthermore, they facilitate discussions, harmonize standards, and promote the development of \gls{satcom} technologies to ensure efficient spectrum usage and equitable access to satellite services across member countries. 
Regional agreements are significant in aligning national interests and obligations with international regulations, enabling smoother coordination for satellite operations across multiple national boundaries.
These frameworks set by global and national authorities ensure that the growing demand for satellite services is managed efficiently and that spectrum resources are used responsibly across different geographical regions.

\section{Dynamic Spectrum Management for Cognitive Satellite Communications} 
\label{sec: dsm for cogsat}

% In general settings, this spectrum is not shared with other network operators. 
% The traditional method of exclusive spectrum allocation is rapidly exhausting the limited available spectrum, making it increasingly difficult for new service providers to enter the market.
% Further, the competition for the wireless spectrum has made it an expensive commodity, making the products delivered over the spectrum more expensive.  
\gls{dsm} has been widely recognized as a promising solution to address spectrum scarcity and has been extensively studied in the literature.
\gls{dsm} is primarily achieved through \gls{cr} and vice versa, thus there exists a huge overlap between the two technologies \cite{liang2020dynamic}.
Keeping that in mind, in this section, we explore \gls{osa}, \gls{csa}, \gls{ss}, and database techniques for \gls{satcom} along with an extensive evaluation of satellite network architectures leading to \gls{cogsat} networks. 
In addition, we categorize literature on \gls{dsm} for \gls{satcom} based on core functionalities and \gls{dsm} techniques, as presented in Table \ref{tab:cr_categorization}.

\begin{table*}[!t]
\centering
\scriptsize
\caption{DSM in SatCom: State-of-the-Art}
\label{tab:cr_categorization}
\begin{tabular}
{|>{\raggedright\arraybackslash}p{0.75cm}
 |>{\raggedright\arraybackslash}p{1.5cm}
 |>{\centering\arraybackslash}p{0.75cm}
 |>{\centering\arraybackslash}p{0.75cm}
 |>{\centering\arraybackslash}p{0.75cm}
 |>{\centering\arraybackslash}p{0.75cm}
 |>{\centering\arraybackslash}p{0.75cm}
 |>{\centering\arraybackslash}p{0.75cm}
 |>{\centering\arraybackslash}p{0.75cm}
 |>{\centering\arraybackslash}p{0.75cm}
 |>{\centering\arraybackslash}p{0.75cm}
 |>{\raggedright\arraybackslash}p{4.25cm}|}
\hline
\multirow{2}{*}{\scriptsize Ref.} & 
\multirow{2}{*}{Network Setup} & 
\multicolumn{4}{>{\centering\arraybackslash}p{3cm}|}{Core Functionalities} & 
\multicolumn{5}{>{\centering\arraybackslash}p{3.75cm}|}{\gls{dsm} Technique} & 
\multirow{2}{*}{Key Contributions} \\ \cline{3-11}
& & \gls{osa} & \gls{csa} & \gls{ss} & \tiny Database/\gls{rem} & \tiny Frequency Reuse & \tiny Power Allocation & \tiny Beam Pointing & \tiny Beam Hopping & \tiny Beam Forming & \\ \hline

\cite{gu2021dynamic} & \gls{geo}-\gls{leo} & - & $\checkmark$ & - & - & $\checkmark$ & $\checkmark$ & - & - & - & Cooperative service method to address the co-linear interference in \gls{geo}-\gls{leo} coexisting network. \\ \hline
\cite{bohai} & \gls{geo}-\gls{leo} & - & $\checkmark$ & - & - & $\checkmark$ & - & - & - & - & \gls{dsa} approach amid \gls{geo}-\gls{leo} system interference. \\ \hline
\cite{yixuan2022underlay} & Satellite-Terrestrial & - & - & $\checkmark$ & - & - & - & - & - & - & Outage performance analysis of terrestrial users under the interference temperature constraint. \\ \hline
\cite{benedetto2021unauthorized} & \gls{geo}-non \gls{geo} & - & - & $\checkmark$ & $\checkmark$ & - & - & - & - & - & A higher order moments-based \gls{ss} approach for detecting unauthorized users. \\ \hline
\cite{ren2021novel} & \gls{geo}-\gls{leo} & - & - & - & - & $\checkmark$ & - & - & - & - & Improved algorithms for frequency reuse in satellite communication. \\ \hline
\cite{yao2023green} & Satellite-Terrestrial & $\checkmark$ & $\checkmark$ & $\checkmark$ & - & - & $\checkmark$ & - & - & - & A two-way-really aided model and a novel power allocation scheme for CSTN. \\ \hline
\cite{feng2023radio} & Satellite-UAV-Terrestrial & - & $\checkmark$ & - & $\checkmark$ & - & - & - & - & - & 6G-NTN network integration approach with coordinated spectrum sharing among satellite-UAV platforms to enhance coverage. \\ \hline
\cite{zhang2021intelligent} & Satellite-Terrestrial & - & - & $\checkmark$ & $\checkmark$ & - & - & - & - & - & An analytical framework for a cloud-based satellite-terrestrial integrated network to improve spectrum utilization. \\ \hline
\cite{zhao2024joint} & \gls{geo}-\gls{leo} & - & - & - & - & - & $\checkmark$ & $\checkmark$ & - & - & An optimization framework for spectrum sharing and interference minimizing in a \gls{geo}-\gls{leo} coexisting environment. A joint model-based and model-free \gls{drl} framework for the proposed framework. \\ \hline
\cite{weththasinghe2023band} & \gls{geo}-\gls{leo} & - & - & $\checkmark$ & - & $\checkmark$ & - & - & - & - & A measu\gls{rem}ent apparatus for frequency reuse opportunities in L-band. A spectrum analysis using Inmarsat data. \\ \hline
\cite{weththasinghe2024optimising} & \gls{leo} - Terrestrial & - & - & - & - & $\checkmark$ & - & - & - & - & Investigation of multiple frequency reuse schemes and beam size optimization approach for \glspl{leo} in multi-beam 6G-\gls{leo} integrated network \\ \hline
\cite{weththasinghe2024cognitive} & \gls{geo}-\gls{leo} & - & - & $\checkmark$ & - & - & - & - & - & - & \gls{ss} platform on \gls{leo} satellites to investigate multiple \gls{geo} spot beam detections. \\ \hline
\cite{li2023joint} & Satellite-Terrestrial & - & - & - & - & - & - & - & $\checkmark$ & - & Beam hopping and adaptive dynamic multiple access scheme to optimize beam scheduling. \\ \hline
\cite{li2023pattern} & Satellite-Terrestrial & - & - & - & - & - & - & $\checkmark$ & $\checkmark$ & - & \gls{qos} guaranteed pattern design and power management scheme for \gls{leo}-Terrestrial beam hopping network setup. \\ \hline
\cite{lin2020robust} & Satellite-Terrestrial & - & - & - & - & - & - & - & - & $\checkmark$ & A secure beamforming approach millimeter wave band sharing satellite-terrestrial network \\ \hline
\end{tabular}
\end{table*}

\subsection{Opportunistic Spectrum Access}

\begin{comment}
\begin{figure}[ht]
    \centering
    \includegraphics[width=0.2345\textwidth]{Images/SpectrumHoles.jpg}
    \caption{Utilization of Spectrum Holes/White Space}
    \label{fig:SpectrumHoles}
\end{figure}
\end{comment}

In \gls{dsm} settings, \gls{pu}s are the privileged user group, as the name implies, and the service provider/network operator has priority over the wireless transmission frequency, as they have bought the rights from a regulatory authority. 
\gls{su}s intend to communicate on the same frequency as the \gls{pu}s with minimal interference between the two systems. 
The literature identifies \gls{pu}s as licensed users and \gls{su}s as non-licensed users, which is a debatable fact, as \gls{su}s are also a responsible user group utilizing/sharing a specific frequency band and should also be licensed and recognized by a regulatory authority. 
This spectrum access policy, where \gls{su}s can transmit without a dedicated frequency band, is known as \gls{dsa}. 
According to \gls{su}s' spectrum access, \gls{dsa} is categorized into two models: \gls{osa} \textit{i.e.,} spectrum overlay or interweave paradigm and \gls{csa}, also referred to as spectrum underlay. 
Table \ref{tab:osa_and_csa} provides a high-level comparison of \gls{osa} and \gls{csa} approaches \cite{liang2020dynamic}. 
\gls{osa} in \gls{cogsat} networks leverages the concept of dynamic utilization of spectrum holes or the portions of the frequency spectrum that are temporarily unoccupied by \gls{pu}s, and allows \gls{su}s to transmit in the identified gaps. 
% The concept of spectrum holes is illustrated in Fig. \ref{fig:SpectrumHoles}. 
% This approach is pivotal in addressing the spectrum scarcity caused by the traditional static allocation~(exclusive licensing) of the spectrum. 
In the broader context of \gls{satcom}, where spectrum is often spatially and temporally underutilized due to the orbital movements of non-\gls{geo} satellites, \gls{osa} enables \gls{su}s to access these unused frequency bands opportunistically, as the name implies \cite{gu2021dynamic}. 
This is enabled through real-time \gls{ss} and geo-location databases, which are discussed in the latter part of this section.  

Unique challenges in \gls{satcom} environments affect the implementation of \gls{osa}, thus the realization of \gls{cogsat} systems. 
Therefore, advanced techniques empowered through wideband \gls{ss} enabled through \gls{sdr}s are critical for detecting spectrum holes across extensive frequency bands. 
Even though \gls{osa} policies have been discussed in the context of terrestrial networks \cite{santivanez2006opportunistic}, it is yet to be fully evaluated and discussed for \gls{satcom} networks. 
This highlights the importance of policy agility, the ability to dynamically adapt to varying regulatory and spectrum usage policies, as it is essential for \gls{osa} in a global satellite context. 
\gls{osa}-driven \gls{cogsat} realization demands the integration of machine-readable policy frameworks that can be updated in real time, enabling seamless operation across different geopolitical regions and spectrum environments.

The primary commercial benefit of \gls{osa} in \gls{cogsat} is the spectral efficiency improvement that \gls{satcom} operators can obtain through successful implementations. 
Such approaches are evaluated extensively in the literature, a \gls{osa} approach for agricultural sensor network deployment is discussed in \cite{bayrakdar2020employing}, while the work in \cite{patil2024comprehensive} discussed \gls{osa} in terrestrial mobile networks. 
Furthermore, a \gls{ml} deployment strategy for \gls{osa} is evaluated in \cite{zhu2020machine}. 
In the broader context, \gls{osa} represents the majority of \gls{cr} concepts evaluated in the state-of-the-art, as it enables the dynamic access of underutilized spectrum with non or minimal interference to \gls{pu}s. 
Moreover, \gls{osa} fosters innovation in \gls{satcom} services, such as broadband internet, \gls{gps}, and remote sensing, by allowing the coexistence of multiple communication systems within the same spectral resources. 

\begin{table}[ht]
\vspace{-2mm}
\caption{{Comparison of OSA and CSA DSM schemes.}}
\begin{center}
\resizebox{\columnwidth}{!}
{
\Large % Adjusts the font size for the entire table
\begin{tabular}{|p{3cm}|p{5cm}|p{5cm}|}
\hline
 & \gls{osa} & \gls{csa} \\
\hline
SU status & On and Off & Always On \\ \hline
Environment awareness  & geo-location database, \gls{ss} & Channel estimations, interference prediction \\ \hline
PU precedence techniques  & Terminate SU transmission through PU detection & Interference control through performance margins \\ \hline
\end{tabular}
}
\end{center}
\label{tab:osa_and_csa}
\vspace{-3mm}
\end{table}

\subsection{Concurrent Spectrum Access}

\begin{figure}[ht]
    \centering
    \includegraphics[width=0.6\columnwidth]{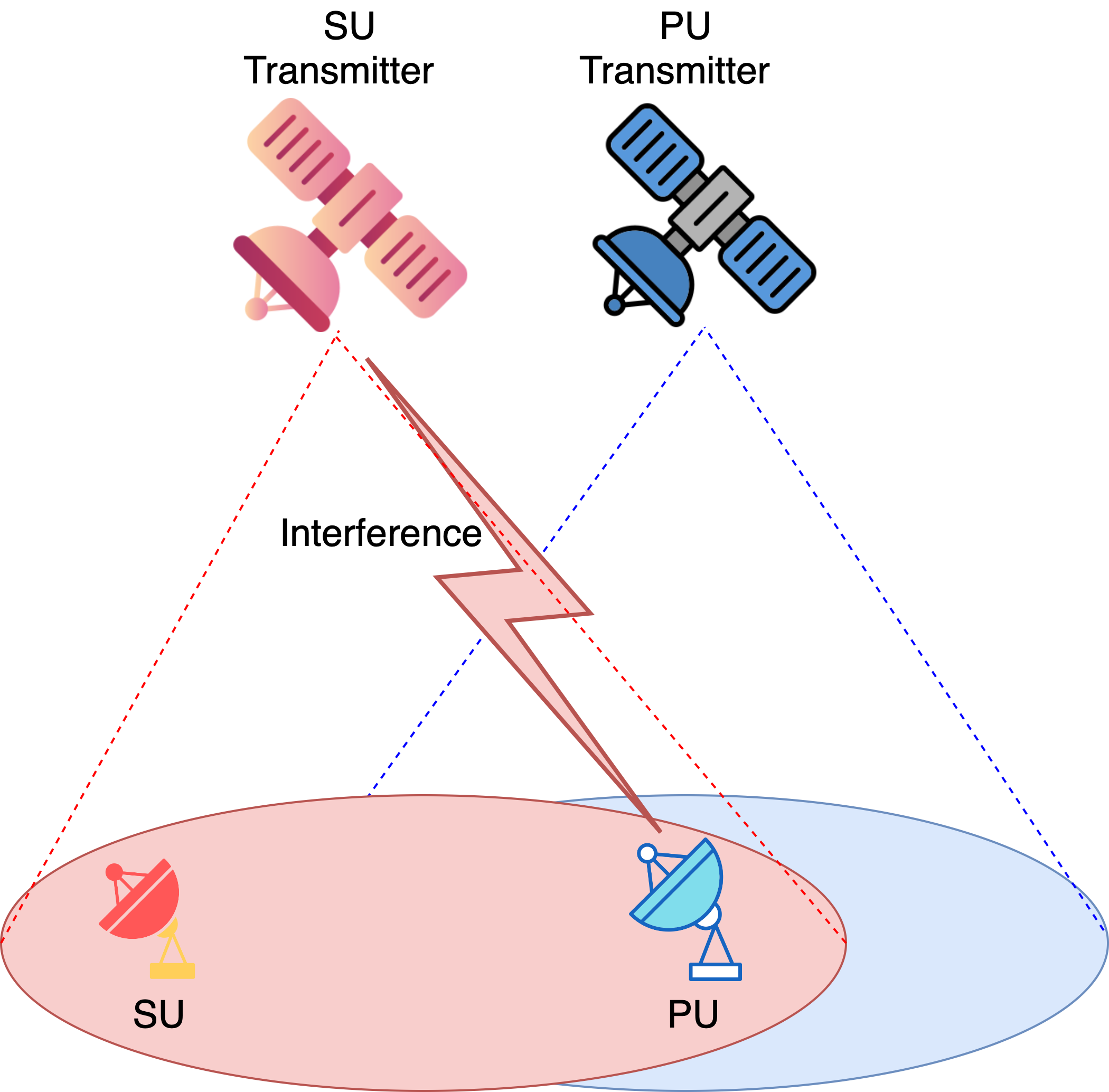}
    \caption{Concurrent spectrum sharing between PU and SUs in SAT environment.}
    \label{img:csa}
\end{figure}

\gls{csa} is another pivotal technique within \gls{cogsat} networks to achieve efficient spectrum utilization. 
Unlike \gls{osa}, where \gls{su}s transmit only when the \gls{pu}s are idle, \gls{csa} enables \gls{su}s to transmit concurrently with \gls{pu}s on the same frequency band. 
This is achieved by ensuring that the interference generated by the \gls{su} transmitter at the primary receiver remains below a tolerable threshold, known as the interference temperature.
This approach allows for continuous \gls{su} transmission, eliminates the need for constant \gls{ss}, and ultimately supports higher spectrum reuse, especially in dense traffic environments such as satellite hotspots.
A typical \gls{csa} model considering \gls{cogsat} environment is illustrated in Fig. \ref{img:csa}.  
The figure further shows how the \gls{su} transmitter inevitably generates interference to the \gls{pu}, in frequency reuse scenarios, if the \gls{pu} is within its transmission range. 
Therefore, to realize \gls{csa}, the \gls{su} transmitter has to deploy cognitive approaches to predict and minimize the interference at the \gls{pu} to an acceptable level, so that the \gls{pu} can achieve the required \gls{qos} level in its transmission.

% In the context of \gls{cogsat} networks, the adoption of \gls{csa} is particularly advanta\gls{geo}us due to the unique characteristics of satellite communication, such as wide-area coverage, long propagation delays, and rigid infrastructure. 
% Here, the SU transmitters, which can be satellite or ground terminals, must be equipped with CSI, particularly cross-channel CSI, to effectively control the interference introduced to the PU receivers. 
% Depending on the nature of the primary service, the interference temperature can be either peak or average interference power constraints. 
% The peak constraint ensures protection at all times, suitable for real-time satellite services, while the average constraint provides more flexibility for non-critical transmissions.

\gls{csa} approaches are typically developed considering multi-user scenarios such as terrestrial cellular networks and satellite networks. 
Therefore, leveraging multi-user diversity to enhance the overall performance of the \gls{su} system by prioritizing \gls{su}s that generate minimal interference to \gls{pu}s represents an extension of the \gls{csa} approach, as thoroughly investigated in \cite{yixuan2022underlay, benedetto2021unauthorized}.
Another \gls{csa} approach is to share the spectrum without an interference limitation to \gls{pu}s. 
In these network setups, both primary and secondary users have equal priorities to access the spectrum. License shared access~\cite{chan2015feasibility}, and spectrum sharing in unlicensed bands explored in~\cite{etkin2007spectrum} are the two main avenues explored.

It is important to note that secondary system throughput and the \gls{qos} guarantees for the primary system are inherently conflicting objectives.
Therefore, making optimal spectrum allocation decisions is vital to reach a balance between the two networks. 
\gls{csa} relies on \gls{csi} estimation methods and interference prediction to share the spectrum with \gls{su}s while maintaining satisfactory service quality for \gls{pu}s. 
The secondary system can make improved decisions when it has access to the primary system \gls{csi} and other critical network information. 
In \cite{SMT2, bohai}, the authors proposed network architectures where primary and secondary satellite systems share network information and user locations to improve spectrum allocation decisions in the secondary system, minimizing the interference to \gls{pu}s.

\subsection{Spectrum Sensing}

% Otherwise known as interwave or interference avoidance, refers to SUs communicating leveraging spectrum holes, which are unused frequencies in the spatial, temporal, and frequency domains—while ensuring no interference with PUs \cite{sharma2012satellite}. 
% The productivity of SS can be improved through exchanging information between neighbours \cite{cabric2004implementation}.  
\gls{ss} in \gls{cr} networks refers to the technique by which \gls{su}s detect and access unused spectrum bands by monitoring the radio environment without interfering with \gls{pu}s. 
Prominent \gls{ss} methods explored used in \gls{cr} networks are as follows:
\begin{itemize}
    \item Energy Detection:  Measures the energy level of the received signal and compares it with a predefined threshold to measure the channel occupancy \cite{ren2021novel}.
    \item Matched Filter Detection: Uses a matched filter to maximize the \gls{snr} for detecting a known signal. Requires prior knowledge of the PU's signal \cite{sharma2012satellite}.
    \item Cyclostationary Feature Detection: Exploits the cyclostationary properties of signals, such as periodicity in their statistics (e.g., mean and autocorrelation) \cite{ngo2023machine}.
    \item Waveform-Based Sensing: Detects known patterns or pilot signals embedded in the primary user transmission \cite{nasser2021spectrum}.
    \item Radio Identification-Based Sensing: Required prior knowledge of \gls{pu} transmission signal properties. Identifies the specific characteristics of the primary user’s signal to determine spectrum usage \cite{yucek2009survey}.
    \item \gls{css}: \gls{cr}s share their \gls{ss} information to improve detection accuracy and mitigate the effects of fading and shadowing \cite{yao2023green}. 
    \item Compressive Sensing: utilizes the sparsity of spectrum occupancy to reconstruct signals using fewer samples than traditional methods \cite{heimerl2024compressed}.
    \item Eigenvalue-Based Detection: Uses the eigenvalues of the covariance matrix of received signals to detect the presence of PUs \cite{chaurasiya2019fast}.
    \item Machine Learning-Based Sensing: Applies \gls{ml} to classify or predict spectrum occupancy based on training data \cite{zheng2020spectrum}.
    \item Hybrid Methods: Combines two or more SS techniques to improve performance \cite{ngo2023machine}.
\end{itemize}
However, the effectiveness of \gls{ss} methods in \gls{satcom} networks is challenged by multi-path fading, large-scale shadowing, and the high variability of satellite transmission channels. 
\gls{css} has emerged as a solution to overcome these limitations by allowing multiple \gls{su}s to exchange sensing information, enhancing detection accuracy, and mitigating the hidden \gls{pu} problem. 
However, unlike terrestrial networks, where \gls{cr} users may collaborate extensively, security and data integrity, along with the distributed nature of the \gls{satcom} network architecture, limit collaboration among \gls{satcom} networks, creating a barrier for \gls{css}. 
Modern advancements exploit additional degrees of freedom, combining two or multiple \gls{ss} techniques, leading to hybrid approaches. 
Furthermore, leveraging \gls{ml} techniques brings an additional dimension to the table with improved prediction capabilities with past data and experience-driven approaches. 
These novel methodologies not only improve the detection accuracy but also minimize sensing time, ensuring faster adaptation to the dynamic satellite environment.
Practical implementations of \gls{ss} in \gls{cogsat} radios must account for unique satellite-specific challenges and characteristics. 
For instance, the long transmission distances often result in low received signal strength for ground users in \gls{satcom} networks, which can inadvertently lead to lower \gls{sinr} values considering the interferences and noise component. 
To counter this challenge, \gls{satcom} systems require low \gls{sinr} sensing capabilities, with thresholds extending as low as -20 dB, as specified in standards like \gls{ieee} 802.22 \cite{cordeiro2005ieee}.

\subsection{Database Technique}

\begin{figure}[ht]
    \centering    \includegraphics[width=0.4\textwidth]{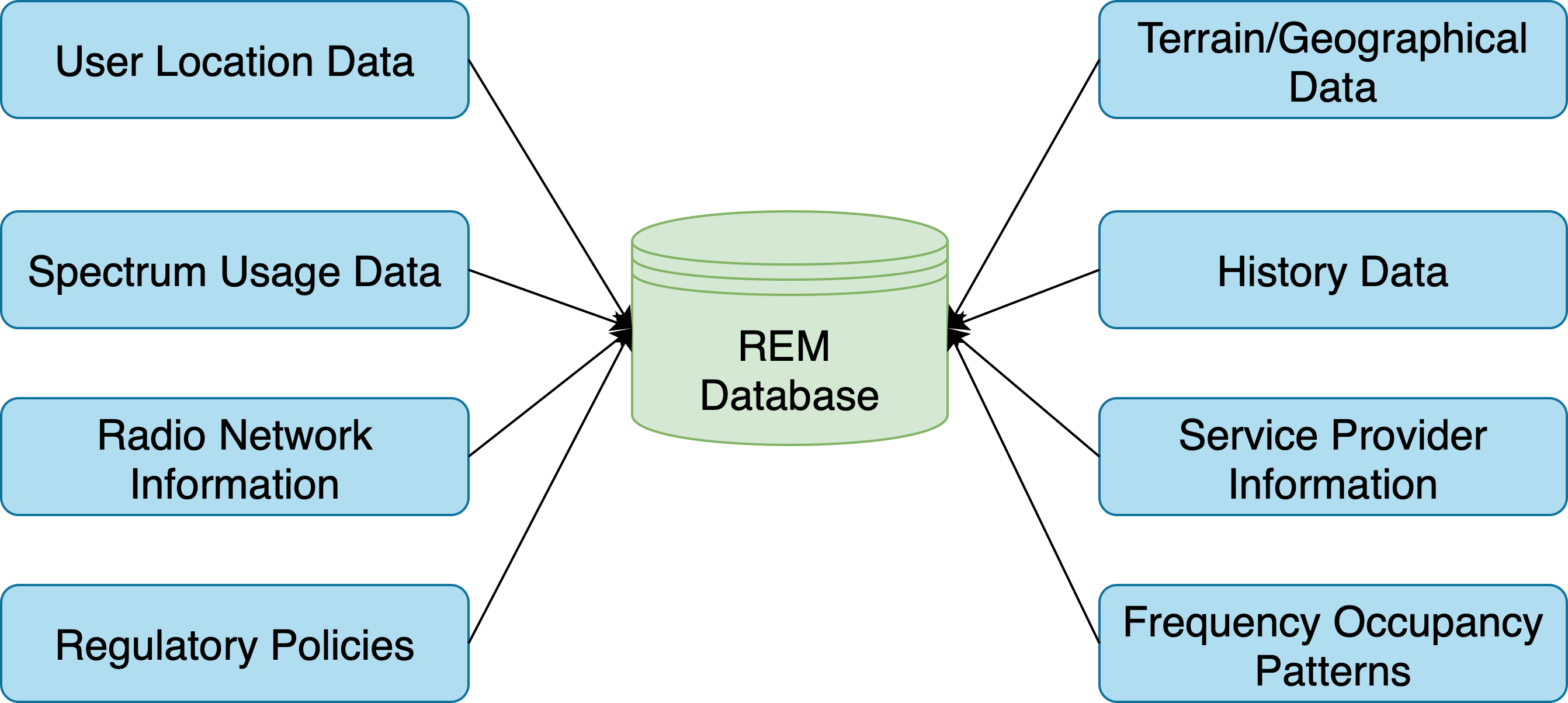}
    \caption{REM database model.}
    \label{fig:REM}
\end{figure}

Database techniques, particularly those involving \gls{rem}s, are pivotal in enabling \gls{dsm} for \gls{cogsat} networks. 
\gls{rem}s function as comprehensive databases containing critical environmental information for \gls{cr} networks, including frequency channel allocations, \gls{rssi}, interference levels, geo-located user activities, and regulatory policies \cite{sayrac2013cognitive, zhao2007applying}, Fig. \ref{fig:REM} further illustrates the data accumulated in a \gls{rem}. 
\gls{rem}s enable environment-aware radio resource management in \gls{cogsat} networks by leveraging direct observations and accumulated network data to construct a precise map of the radio environment. 
Therefore, in the context of \gls{cogsat} radios, \gls{rem}s are invaluable tools for tasks such as predicting \gls{pu} frequency usage patterns, enabling \gls{su}s to identify and access white spaces, and facilitating spectrum sharing in both licensed and unlicensed bands.
Thus, minimizing interference towards \gls{pu} and \gls{su}s while improving overall spectral efficiency. 
Key database deployment approaches for \gls{cr} are as follows:
\begin{itemize}
    \item Centralized Database: A centralized database manages all the data related to spectrum availability. A central authority could manage and could be deployed leveraging on-premises or cloud deployment options \cite{vizziello2011system}. 
    \item Distributed  Database: Spectrum information is distributed across multiple nodes, avoiding reliance on a central database and minimizing communication latency, thus optimizing the database access process \cite{ezzati2018distributed}.
    \item Hierarchical Database: Combines centralized and distributed approaches, leading to a hierarchy of databases \cite{zhao2006overhead}. 
    \item Blockchain-based database: Leverage blockchain methods leading towards secure and decentralized spectrum management databases \cite{chen2022blockchain}.
\end{itemize}
In addition, hybrid approaches with combinations of multiple database management techniques can be utilized to enhance the overall efficiency.

\gls{rem}s can be further categorized into direct, indirect, and hybrid approaches based on the data collection and distribution methods, each offering unique insights into the radio environment \cite{pesko2014radio}. 
Direct methods rely on real-time measurements, while indirect methods synthesize data through statistical models and historical information, and hybrid methods combine these approaches to enhance accuracy and predictive capabilities. 
These approaches highlight the role of \gls{rem} in fostering adaptive and efficient spectrum management, even in environments where direct radio environment measurements are unavailable.
In addition, at a structural level, \gls{rem}s exist in two main forms, local \gls{rem}s and global \gls{rem}s. 
Local \gls{rem}s, synchronized with their global counterparts, facilitate individual users or user groups with tailored \gls{cr} environment insights. 
This hierarchical structure, which combines centralized and distributed databases, enhances the decision-making processes of \gls{cr}s, enabling them to learn from accumulated experiences and adapt to changing conditions \cite{zhao2006overhead}. 
By integrating \gls{rem} data, \gls{cr} networks can reduce processing overhead and adaptation times, making large-scale deployments such as \gls{cogsat} networks more cost-effective in \gls{dsm}. 
Furthermore, \gls{rem}s enable advanced network functionalities such as situation awareness and multi-domain knowledge sharing, further enhancing the \gls{cr} network's overall intelligence and adaptability \cite{4221484}.

The practical deployment of \gls{rem} is exemplified in systems like the \gls{wran} proposed in \cite{zhao2007applying}, where sensing and measuring data are leveraged for realization.
Using geo-location data and querying centralized \gls{rem}-enabled databases, \gls{wran} systems dynamically identify available spectrum resources, including operating parameters like channel availability, center frequencies, and power levels, ensuring \gls{pu} protection while maximizing \gls{su} performance. 
Advanced \gls{rem} architectures, enhanced through spatial statistical modeling and topology engines, further refine spectrum management capabilities. 
These innovations make database-driven approaches a cornerstone of \gls{dsm} techniques, enabling \gls{cogsat} radios to navigate complex and dynamic spectrum environments with enhanced efficiency and reliability.

\subsection{Network architectures for Cognitive Satellites}

\gls{cogsat} networks can be broadly categorized into;
\begin{itemize}
    \item Integrated \gls{cogsat}, Hybrid \gls{cogsat}, \gls{cstn}- Coexistence of satellite and terrestrial network sharing the same spectrum 
    \item Dual \gls{cogsat} - Inter satellite system spectrum sharing,
\end{itemize} 
based on the network architecture \cite{SMT2}.
% Different CR techniques should be utilized in these scenarios as the network setups have drastic differences. 

\subsubsection{Integrated Cognitive Satellite Networks}

\gls{osa} in integrated \gls{cogsat} or \gls{cstn} can be categorized based on the spectrum presidence, the first one being the satellite network taking priority over the terrestrial networks, where terrestrial users access the transmission frequency with minimal interference to the primary satellite users. 
A multi-beam satellite network operating as the primary system while sharing the frequency with randomly distributed terrestrial \gls{bs} is discussed and evaluated in \cite{kolawole2017performance}, where a \gls{bs} thinning process is proposed to minimize the interference below a predefined primary system requirement. 
A time-splitting spectrum sharing approach is investigated in \cite{9187830} where the primary satellite network shares the spectrum with the secondary terrestrial network. 
Similar network models have been investigated along different aspects in \cite{zhu2021integrated, 9628097} and the references therein.  
The second approach is the inverse of the first scenario, where the terrestrial network gets the priority in accessing the dedicated frequency band, and the satellite network shares the terrestrial network frequency with minimal interference to the primary terrestrial users. 
Therefore, a transmission power and carrier allocation methodology is proposed in \cite{7336495} for an integrated \gls{cogsat} network where satellites exploit the microwave frequency band allocated to terrestrial networks. 

\gls{csa} is also a possibility in integrated \gls{cogsat}, where both satellite and terrestrial networks utilize the shared spectrum simultaneously maintaining a maximum interference threshold for \gls{pu}s \cite{8786872}. 
Performance of such networks, concerning interference power constraints imposed by terrestrial communication regulations, is evaluated with regard to bit error rate and network outage in \cite{ruan2018performance}.  
An underlay \gls{cstn} where satellites operate in microwave frequency allocated to terrestrial use is proposed and evaluated in \cite{ruan2017effective}, and considering statistical delay \gls{qos} requirements, satellite network effective capacity is investigated under terrestrial imposed interference power limitations.
\gls{noma} along the direction of spectral efficiency \cite{le2021enabling} and signal relays in cooperative \gls{noma} \cite{singh2020underlay, 8949359} is investigated extensively for underlay \gls{cstn} in the literature. 

\begin{figure}[h]
    \centering
    \includegraphics[width =\columnwidth]{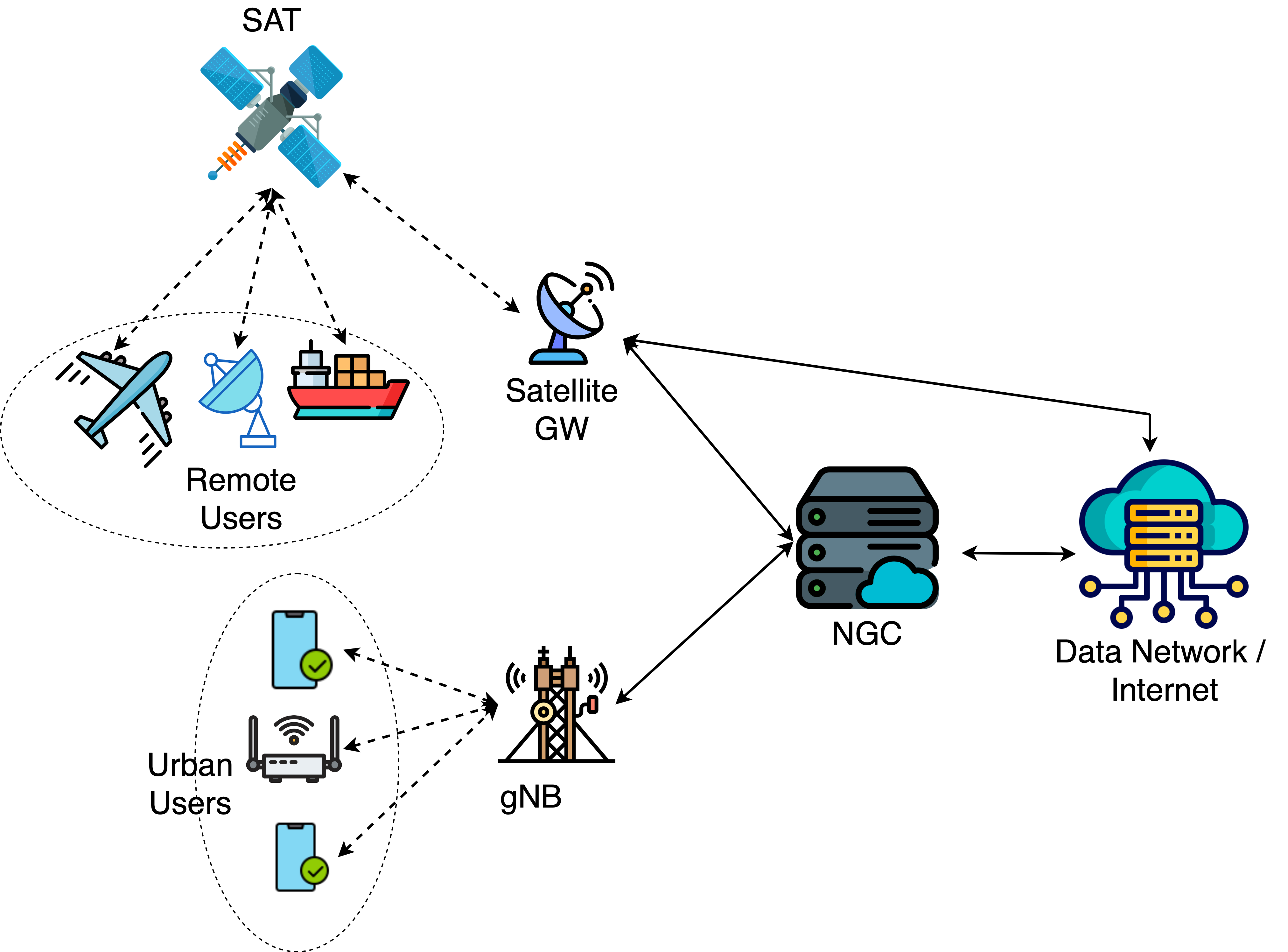}
    \caption{CI-CSTN network architecture.}
    \label{fig:ci_cstn}
    \vspace{-5mm}
\end{figure}

Another integrated \gls{cogsat} approach discussed in the literature is \gls{cicstn}, which is an additional step towards realizing satellite-terrestrial networks \cite{7336495}.
In \gls{cicstn}, the \gls{satcom} network connects remote terrestrial network users, while the terrestrial network connects the rest. 
These networks enable the utilization of a common frequency band with minimal interference between the satellite and terrestrial networks. 
Thus, improving the spectral utilization efficiency with minimal inter-network interference. 
A \gls{cicstn} network architecture is illustrated in Fig. \ref{fig:ci_cstn}. 
Considering the current direction of satellite-terrestrial network integrations, \gls{cicstn} are the most likely to be realized in a large-scale network. 

\subsubsection{Dual Cognitive Satellite Networks}

\begin{figure}[h]
    \centering
    \includegraphics[width = \columnwidth]{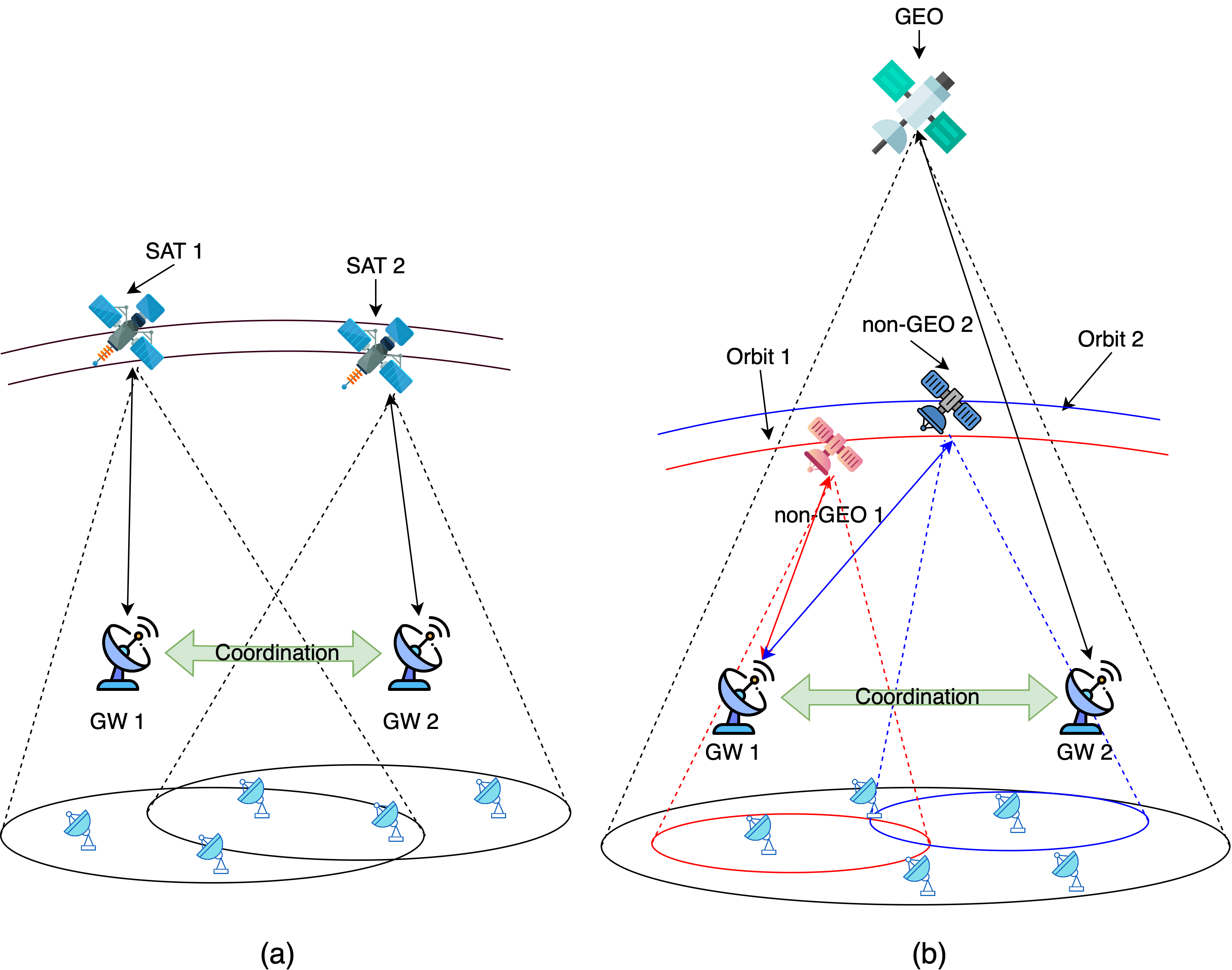}
    \caption{Dual CogSat scenarios (a) Same type of satellite coexisting. (b) GEO and non-GEO satellites coexisting.}
    \label{fig:dual_cogsat}
\end{figure}

Refers to the scenario where two satellite systems operate simultaneously over a coverage area utilizing the same spectrum band \cite{gupta2015cognitive, bohai}. 
Spatial and spectral degrees of freedom are shared between the satellite systems in these networks. 
Based on the network architecture, literature on dual \gls{cogsat} can be mainly categorized into the same type of satellites coexisting networks and \gls{geo} and non-\gls{geo} satellite coexisting networks, as illustrated in Fig. \ref{fig:dual_cogsat}. 
The same kind of satellites can be either \gls{geo}, \gls{meo}, or \gls{leo}, sharing the frequency to serve a common area of interest. 
They can be from the same constellation or different constellations, but should utilize \gls{cr} techniques in their deployments. 
Due to the growing demand for satellite-based services, multiple satellites can be deployed in close proximity, thus creating overlapping coverage areas. 
The fixed satellite services deployed to serve hot orbital zones like 13E and 19E with orthogonal frequency plans are conventional use cases for such networks \cite{gupta2015cognitive}. 
Further, these satellite networks can leverage both mono-beam and multi-beam technologies to serve the users depending on the use case and capacity requirements. 
A study exploring the coexistence of multi-beam \gls{geo} satellites was done in \cite{nguyen2017cognitive}, in which the authors propose a cognitive beam-forming approach to mitigate the uplink co-channel interference. 

In \gls{geo} and non-\gls{geo} satellite coexistence networks, in-line interference is an additional component to the interference generated from the co-located satellites. 
Further, orbital relative motions of non-\gls{geo} satellites add further complexities to these networks. 
To this end, Skybridge and Teledesic \gls{leo} satellite systems proposed to reuse the \gls{geo} frequency band for their transmission. 
Ideally, Skybridge \glspl{leo} proposed to terminate transmission within a certain distance from the equator to mitigate interference to \gls{geo} users \cite{ashford2004non}. 
On the other hand, Teledesic planned to terminate transmission when the satellite coverage footprint intersects with the equator. 
Earth terminals would only transmit with the Teledesic \gls{leo} satellites if their latitude is northerly within the sub-satellite point in the northern hemisphere, and the relative sub-satellite point latitude has to be southerly in the southern hemisphere to initiate a successful transmission \cite{stuart1998frequency}. 
However, neither of these cognitive frequency reuse approaches are materialized \cite{gupta2015cognitive}.

\subsection{Dynamic Spectrum Management Techniques}

% As discussed above \gls{dsm} in \gls{cogsat} radios is a cornerstone for optimizing spectrum utilization in \gls{satcom} networks. 
% With the emergence of future network architectures integrating \gls{geo}, \gls{meo}, and \gls{leo} satellites, along with the global expansion of mega-\gls{leo}constellations, efficient spectrum allocation has become paramount in this rapidly evolving industry.
Techniques such as frequency reuse, power allocation, beam pointing optimization, and beam hopping are integral in realizing \gls{dsm} in \gls{satcom} systems.
The key objectives of these \gls{dsm} techniques in a \gls{cr} setting are to improve spectral efficiency, maximize the \gls{pu} and \gls{su} system throughput, and minimize or eliminate the \gls{pu} interference while maintaining an operational level of interference for \gls{su}s.

\subsubsection{Frequency Reuse}

This is a widely adopted \gls{dsm} technique where the same frequency band is reused in the non-overlapping coverage area, maximizing spectrum efficiency.
This approach is already in operation in both terrestrial \cite{xie2010frequency} and non-terrestrial \cite{kisseleff2020radio} networks under general (non-cognitive) settings, as the licensed spectral bandwidth for each operator is limited and the user capacity requirements are high. 
For instance, the Ka-band in \gls{hts}s employs a frequency reuse factor of 4 or higher \cite{hasan2016ka}, effectively dividing the available spectrum across a multi-beam environment, while minimizing inter-beam interference.
This approach can be extended for \gls{cr} through opportunistic frequency reuse techniques, as \gls{su} reusing \gls{pu} spectrum with non or minimal interference to the \gls{pu} network \cite{luo2022research}.
In the context of \gls{cogsat} radios, intelligent frequency reuse schemes utilize advanced algorithms to dynamically adjust the reuse pattern based on user density, interference levels, satellite dynamics, and traffic demands \cite{bohai}. 
Moreover, the \gls{3gpp} identifies frequency reuse as key for efficient spectrum utilization in \gls{ntn} starting from Release~17~\cite{giordani2020non}, thus highlighting the importance of this technique going forward with the standardized implementations. 
Through advanced \gls{ss} and \gls{rem} approaches, the frequency reuse technique can be further enhanced to improve the spectral efficiency. 

\subsubsection{Power Allocation}

A critical technique in enabling \gls{cr} networks, which refers to dynamic power adjustment of \gls{su} transmission to optimize spectrum utilization while minimizing interference to \gls{pu}s. 
This efficient power allocation approach ensures the \gls{su} operation with minimal disruptions to the \gls{pu} network, achieving a balance between performance, energy efficiency, and spectrum fairness.
In the context of \gls{cogsat} networks, the power allocation technique ensures efficient distribution of transmission power across different beams, users, and frequency channels while optimizing link quality and spectrum utilization.
Literature has explored several key approaches in realizing power allocation for \gls{cogsat} systems, such as game theory \cite{del2009power}, optimization \cite{ge2021joint}, and \gls{ml}-based \cite{guo2022multi}. 
Furthermore, this technique can be extended to maintain the \gls{qos} levels, as \gls{cr} refers to adapting radio network parameters according to the operational environment. 
For example, power allocation can adapt to rain fade conditions dynamically by increasing power in affected beams \cite{destounis2011dynamic}. 
In multi-beam systems, power allocation-based cognitive mechanisms can be deployed to analyze interference patterns and adapt beam power levels to mitigate inter-beam interference, enhancing overall network capacity \cite{hajipour2020computing}.

\subsubsection{Beam Pointing}

Refers to dynamically adjusting the directions and shape of the radio beam, while maximizing the coverage and minimizing the interference to \gls{pu}s, while improving spectral efficiency. 
This approach is also leveraged to cater to higher user demands, as the existing radio resources can be exhausted under unplanned scenarios. 
In \gls{satcom} radio networks, the beam pointing optimization approach uses advanced algorithms \cite{he2024interference} and \gls{ml} models \cite{zhang2022deep} to predict traffic patterns, thus resulting in beam steering. 
For instance, due to dense deployment and high mobility, \gls{leo} satellite users in mega constellations can face interference issues, such problems can be addressed through situational aware beam pointing optimization approaches \cite{he2024interference}. 
Furthermore, this approach can be extended into \gls{pu} and \gls{su} operating \gls{cogsat} environments to opportunistically utilize frequency bands, generating minimal interference to \gls{pu} users \cite{he2024joint}.

\subsubsection{Beam Hopping}

This technique represents the dynamic allocation of beams to provide coverage for different geographical areas. 
Beam hopping enables service facilitation across multiple regions using a single beam by allocating time slots.
\gls{cogsat} networks, with the integration of \gls{ss} and \gls{rem} techniques, leverage this approach for adaptive resource allocation. 
For example, \gls{leo} satellites can schedule beam arrangements based on dynamic user demands. 
The available spectrum resources are also time-varying due to the dynamics of the satellites, as beam hopping facilitates time slot-based operations.  
This technique can facilitate agile beam allocations under constraints, and such approaches for \gls{cstn} are proposed in \cite{li2023joint, li2023pattern}

\subsubsection{Beam Forming}
Refers to shaping the radiation pattern of the antenna array for concentrated energy transmission toward desired directions and suppressing interference, thus creating directional radiation patterns for a specific user or a region. 
This can be leveraged for \gls{cr} networks, as it enables \gls{su}s connection amid the \gls{pu}s in the radio environment. 
In the context of \gls{cogsat}, advanced beam forming techniques are leveraged to adapt to varying spectrum and traffic demands in real time, thus directing the transmission energy of identified beams toward specific users or regions, maximizing signal quality, and minimizing interference of the overall communication system. 
To realize this technique, literature has explored advanced optimization schemes such as the discretisation and the Taylor expansion combined approach proposed in~\cite{lin2020robust} for \gls{cstn}, and the penalty function-based approach discussed in \cite{zhao2021outage} for \gls{cogsat}-Arial network setup. 
Furthermore, the potential of advanced \gls{ml} methods towards this is also explored in \cite{zhang2022deep}, for \gls{mimo} in \gls{leo} satellite systems. 
Moreover, digital, analogue, and hybrid beam forming approaches are explored towards advancing this technique in \cite{al2022review, rao20215g} and the references within provide a deep dive into those approaches.

\section{Machine Learning in Satellite Communication and Spectrum Management}
\label{sec: ML in Sat}
This section explores \gls{ml} and \gls{ai} methods leveraged in \gls{ss}, spectrum allocation, interference mitigation and resource management. 
We further discuss \gls{ml} model training and operational resilience, while extensively categorizing the state of the art \gls{ml} methods on satellite spectrum management in Table \ref{tab:ml_cr}. 

\begin{table*}[!t]
\centering
\scriptsize
\caption{Literature on leveraging ML for Intelligent Spectrum Management.}
\label{tab:ml_cr}
\begin{tabular}
{|>{\centering\arraybackslash}p{0.75cm}
 |>{\centering\arraybackslash}p{0.85cm}
 |>{\centering\arraybackslash}p{0.85cm}
 |>{\centering\arraybackslash}p{0.85cm}
 |>{\centering\arraybackslash}p{0.85cm}
 |>{\centering\arraybackslash}p{1.2cm}
 |>{\centering\arraybackslash}p{0.7cm}
 |>{\centering\arraybackslash}p{0.7cm}
 |>{\centering\arraybackslash}p{0.7cm}
 |>{\centering\arraybackslash}p{0.7cm}
 |>{\centering\arraybackslash}p{5cm}|}

\hline
\multirow{2}{*}{Ref.} 
& \multicolumn{4}{>{\centering\arraybackslash}p{4cm}|}{\gls{ml} Categorization \& Algorithms} 
& \multirow{2}{*}{\makecell{Network\\Setting}} 
& \multicolumn{4}{>{\centering\arraybackslash}p{4cm}|}{Focus Area} 
& \multirow{2}{*}{Key Contributions} \\ 
\cline{2-5} \cline{7-10}
& \centering \gls{sl} 
& \centering \gls{usl}  
& \centering \gls{rl}  
& \centering  Other 
&  
& \centering  \gls{ss} 
& \centering\tiny Interference Detection 
& \centering\tiny Spectrum Allocation 
& \centering\tiny Resource Management 
& \\ 
\hline

\cite{9910435} & \acrshort{cnn}-\acrshort{lstm} & - & -  & - & \gls{geo} & $\checkmark$ & - & - & - & Realistic data collected from Tiantong-1 \gls{geo} satellite is used as training data. \\ \hline
\cite{10000832} & \acrshort{lstm} & - & -  & - & \gls{geo} & $\checkmark$ & - & - & - & Realistic data collected from Tiantong-1 \gls{geo} satellite is used as training data. \\ \hline
\cite{9387707} & \acrshort{cnn} & - & -  & - & Satellite-Terrestrial & $\checkmark$ & - & - & - & Blind threshold algorithm eliminating the impact of noise uncertainty. Performance is evaluated against simulated and real-world data. \\ \hline
\cite{ngo2023machine} & SR, LR, DT, SVM & - & -  & - & \gls{geo}-\gls{leo} & $\checkmark$ & - & - & - & Multiple supervised learning approaches evaluated with cyclostationary feature detection. \\ \hline
\cite{9293140} & \acrshort{cnn}-BiLSTM & - & -  & - & \gls{geo}-\gls{leo} & $\checkmark$ & - & - & - & Spectrum prediction evaluated under \gls{leo} using shared spectrum. Results were evaluated against several supervised learning algorithms. \\ \hline
\cite{ngo2023multi} & - & - & \acrshort{madrl} \acrshort{dqn}  & - & Satellite-Terrestrial & $\checkmark$ & - & - & - & \acrshort{madrl} approach for reconfigurable intelligent surfaces assisted cognitive satellite-terrestrial networks. \\ \hline
\cite{ferreira2018multiobjective} & - & - & \acrshort{dqn}  & - & Satellite-HAP & - & - & - & $\checkmark$ & Proposed approach is a part of the PoC for International Space Station's CR engine. \\ \hline
\cite{liu2021gnss} & Standard NN \& \acrshort{cnn} & -  & - & - & Satellite-Terrestrial & - & $\checkmark$ & - & - & Explore GNSS interference events at airports that can affect airplane landings. \\ \hline
\cite{saifaldawlaconvolutional} & - & CAE & -  & - & \gls{geo}-N\gls{geo} & - & $\checkmark$ & - & - & CAE-based method for non-\gls{geo} satellite interference detection at \gls{geo} users. \\ \hline
\cite{he2024collaborative} & - & - & \acrshort{ppo}   & - & \gls{geo}-\gls{leo} & - & $\checkmark$ & - & $\checkmark$ & \gls{drl} approach to resolve co-channel interference in \gls{geo}-\gls{leo} coexisting setup. \\ \hline
\cite{9273081} & - & - & \acrshort{dqn}  & - & \gls{ntn} & - & - & $\checkmark$ & - & Spectrum access approach accounting NTN BS dynamics. \\ \hline
\cite{9047860} & - & - & DeepCA  & - & \gls{leo}-\gls{siot} & - & - & $\checkmark$ & - & A novel \gls{drl}-based approach dubbed DeepCA for optimal channel allocation considering energy constraints. \\ \hline
\cite{8372935} & - & - & \acrshort{dqn} & - & \gls{geo}-\gls{siot}  & - & - & $\checkmark$ & - & \gls{dsa} method for multibeam satellite systems leveraging image-like tensors to extract environment information. \\ \hline
\cite{Liu2018} & - & - & \acrshort{dqn}-CNN  & - & \gls{geo}-\gls{siot}  & - & - & $\checkmark$ & - & \gls{dsa} algorithm for multibeam satellite systems. \\ \hline
\cite{9387703} & - & - & -  & AI & Satellite-Terrestrial & - & - & $\checkmark$ & - & \gls{sdn} \& AI integrated approach toward intelligent spectrum management. \\ \hline
\cite{guo2022multi} & - & - & MADDP  & - & Satellite-HAP & - & - & - & $\checkmark$ & Trajectory and power optimization to reduce latency. \\ \hline
\cite{mishra2024traffic} & - & - & -  & RNF & Satellite-Terrestrial & - & - & - & $\checkmark$ & \gls{sdn}-based spectrum sharing/traffic offloading using RNF and feed-forward NNs. \\ \hline
\cite{ML6} & - & - & Actor Critic-\acrshort{dqn} & - & Heterogeneous SAT & - & - & $\checkmark$ & - & \gls{drl} and \gls{madrl} approaches to optimize resource utilization in \gls{sdn}/\gls{nfv}-enabled networks. \\ \hline
\cite{abdu2022deep} & SMDL \& MMDL & - & -  & - & \gls{geo} SAT & - & - & $\checkmark$ & - & SMDL \& MMDL-based accelerated method for bandwidth/power allocation. \\ \hline
\cite{Xin2018} & - & - & \acrshort{cnn}-\acrshort{dqn}  & - & \gls{geo}-\gls{siot} & - & - & - & $\checkmark$ & Image tensor-based state reformation approach for spatial/temporal feature capture. \\ \hline
\cite{Liao2020} & - & - & \gls{madrl} \acrshort{dqn}  & - & \gls{geo}-\gls{siot}  & - & - & $\checkmark$ & $\checkmark$ & Cooperative dynamic \gls{madrl} approach for distributed intelligence. \\ \hline
\cite{ortiz2022machine} & - & - & \acrshort{cnn} \& \acrshort{dqn}  & - & \gls{geo}-\gls{siot}  & - & - & $\checkmark$ & $\checkmark$ & Power, bandwidth, and beam hopping across \gls{sl} and \gls{rl} layers. \\ \hline
\cite{ortiz2022supervised} & LR & - & -  & - & High throughput-\gls{siot} & - & - & $\checkmark$ & - & Power and bandwidth allocation. \\ \hline
\cite{duong2023machine} & \acrshort{dqn} & - & -  & - & Satellite-Terrestrial & - & - & - & $\checkmark$ & Multi-beam approach using \gls{dqn} optimization and game theory. \\ \hline
\cite{saifaldawla2024genai} & - & - & -  & GenAI & \gls{geo}-\gls{leo} & - & $\checkmark$ & - & - & VAE and TrID-based Generative AI to mitigate \gls{leo}-to-\gls{geo} interference. \\ \hline
\cite{zhao2024joint} & - & - & JMB-\gls{ml}  & - & \gls{geo}-\gls{leo} & - & - & - & $\checkmark$ & Joint model-based/model-free \gls{drl} for beam and resource management. \\ \hline
\cite{sadique2025link} & \acrshort{rnn} \& \acrshort{cnn} & - & -  & - & \gls{leo} & - & - & - & $\checkmark$ & Link scheduling in over Riemannian Manifolds \\ \hline
\cite{yang2024csl} & - & - & -  & \acrshort{cnn}, self-attention, \acrshort{lstm}, \& soft fusion & \gls{geo}-\gls{leo} & $\checkmark$ & - & - & - & \acrshort{css} model to improve detection performance\\ \hline
\cite{tang2024dynamic} & - & - & DDQN \& DDPG \gls{madrl}  & - & satellite-terrestrial & - & - & $\checkmark$ & - & Multichannel LEO spectrum sharing framework for terrestrial users\\ \hline

\end{tabular}
\end{table*}

\subsection{Machine Learning for Spectrum Sensing}

% \gls{ss} involves detecting unutilized spectrum bands~(spectrum holes), in dynamic and often unpredictable radio environments. 
% In order to leverage the full potential of spectrum holes, it is essential to predict the availability and unavailability states of the radio frequency channel, thus reducing the risk of interference while improving communication reliability. 
% \gls{ml} algorithms are well-explored and employed for pattern classification, where features extracted from training data are used to train a model to predict the patterns. 
\gls{ss} can be characterized as a binary classification problem that can be solved using supervised and unsupervised \gls{ml} algorithms, in which the classifier has to determine the availability and unavailability of a radio channel of interest. 
Energy and probability vectors can be used as features in \gls{ml} algorithms to predict spectrum hole availability \cite{janu2022machine}. 
\gls{lstm} is a classification of \gls{rnn}, with the capability to effectively learn and remember long-term dependencies in sequential data. 
% Further, it employs specialised gates to control the flow of information to overcome the vanishing gradient problem, thus making it appropriate for time series forecasting. 
A study of \gls{ss} for \gls{satcom}, taking the propagation delay into calculations, is presented in \cite{10000832}. 
The authors highlight that neglecting the propagation delay in satellite links can cause co-frequency interference at ground level, and to mitigate this, they propose a \gls{dl}-based joint \gls{lstm} and autoregressive moving average~(LSTM-ARMA) \gls{ss} scheme.
Reliable \gls{ss} under low \gls{snr} conditions, otherwise identified as \gls{snr}-wall, is a key challenge in \gls{ntn}s.
To this end, a \gls{ss} scheme using a combined convolutional neural network and long short-term memory~(C-CNN-LSTM) to mitigate the effect of low \gls{snr} is proposed in \cite{9910435}.  
Further, a \gls{cnn}-based approach to improve \gls{ss} under low \gls{snr} conditions in space-air-ground integrated networks is evaluated in \cite{9387707}, where the authors derive a likelihood test for \gls{ss} under the Neyman-Pearson lemma using \gls{cnn}.   

In \cite{ngo2023machine}, the authors have explored a cyclostationary feature detection method in the context of dual satellite networks and proposed an \gls{ml} approach to improve the \gls{ss}. 
Further, they have evaluated the performance of the \gls{svm}, Decision Tree, Logistic Regression, and Softmax Regression supervised learning algorithms for dual satellite network \gls{ss}. 
\gls{dqn} is an \gls{rl} algorithm that combines Q-learning, which maximizes the expected future rewards of taking a given action in a given state, of a defined \gls{rl} environment, with deep neural networks. 
\gls{dqn} enables the agent to learn optimal policies directly from high-dimensional sensory inputs. 
In \cite{kumar2022deep}, a \gls{dqn}-based \gls{ss} approach is proposed and evaluated against an energy detection algorithm and a \gls{cnn} algorithm. The authors highlight the performance improvement of the proposed \gls{drl} methodology, even with the quantity of relatively small training data.  
However, the full potential of \gls{drl} against \gls{ss} for \gls{satcom} networks is yet to be fully explored in the literature.

\subsection{Machine Learning for Spectrum Allocation}

Spectrum allocation in satellite networks under \gls{cr} settings itself is a complex problem governed by the temporal, spectral, and frequency characteristics of the environment. 
Conventional spectrum allocation methods often rely on static rules or simplistic models that fail to adapt to the rapidly changing conditions of \gls{satcom} networks.
However, \gls{ml} offers robust solutions to this complex problem, taking multiple factors such as user demand, signal strength, and interference levels into account, and makes real-time spectrum allocation decisions.
Particularly, \gls{drl} algorithms with the inherited ability to interact with the environment and learn from the feedback have shown superior performance in solving multi-dimensional problems similar to spectrum allocation in \gls{satcom} networks.  
The adaptability and predictive capabilities based on historical data and real-time inputs of \gls{drl} algorithms make it a powerful tool for enhancing the spectral efficiency of SAT systems.

A \gls{dca} methodology for multi-beam satellites leveraging \gls{dqn} is proposed in \cite{8372935}, where the authors introduce an image-like tensor to represent the state, thus encapsulating spatial and temporal features of the \gls{satcom} environment.  
\gls{leo} satellites are designed with power constraints to reduce production and deployment costs; thus, power efficiency is the dominant factor in highly dynamic \gls{leo} satellite constellations. 
Therefore, a power-efficient channel allocation approach empowered by \gls{drl} is presented in \cite{9047860} contemplating a satellite and \gls{iot} environment. 
In addition, a \gls{dqn}-driven multi-user access control approach for \gls{ntn} is presented in \cite{9273081}, where the authors propose a methodology to improve the long-term throughput of the ground users by minimizing frequent handovers. 
\gls{dca} methods can be used to minimize co-channel interference in \gls{satcom} systems. 
Therefore, mitigating the flows in \gls{dca} based on beam traffic load and user terminal distribution, an improved \gls{drl}-based \gls{dca} algorithm for multi-beam satellite systems is proposed in \cite{Liu2018}, to minimize service blocking probability.

\subsection{Machine Learning for Interference Mitigation}

In order to facilitate high throughput requirements in modern applications, and to improve spectral efficiency frequency reuse techniques are leveraged, and it is a widespread methodology deployed in almost every wireless \gls{wan}s. 
Frequency sharing and reuse within and between satellite networks generate co-channel and inline interference, which is an identified challenge \cite{SMT2}. 
Interference minimization in satellite networks can be achieved primarily through adjusting parameters such as \gls{eirp}, antenna direction, and frequency planning. 
The capabilities of \gls{ml} techniques facilitate improved solutions in predicting these communication parameter adjustments amid the highly topological and dynamic characteristics of \gls{satcom}.
\gls{ml} models analyze vast amounts of real-time data from satellite sensors, ground stations, and user terminals to detect patterns and anomalies, enabling proactive measures to prevent or reduce interference, thus leading to optimized SAT networks facilitating improved service to the users.

Autoencoders are neural networks designed to learn efficient and compressed data representations by encoding input data into a lower-dimensional latent space, reconstructing the original data from this compressed representation. 
A convolutional autoencoder-based interference detection approach is proposed in \cite{saifaldawlaconvolutional} for \gls{geo} and non-\gls{geo} coexisting satellite networks. 
Further, a Generative AI~(GenAI) methodology for interference management for \gls{geo} frequency sharing with non-\gls{geo} satellites is present and evaluated in \cite{saifaldawla2024genai}. 
A dynamic interference management methodology for \gls{leo} downlink is presented in \cite{yun2023dynamic}, where the authors elaborate on \gls{dqn} performance over other \gls{ml} algorithms on downlink throughput maximization. 
A \gls{cnn} based approach to detect and mitigate interference in the \gls{gnss} was proposed in \cite{liu2021gnss}. 
In contrast, an \gls{lstm} algorithm for interference detection in \gls{satcom} networks is evaluated in \cite{9384473}.
In addition, a collaborative interference avoidance method for \gls{geo}-\gls{leo} coexisting satellite systems leveraging the \gls{ppo} algorithm is proposed in \cite{he2024collaborative}.

\subsection{Machine Learning  for Resource Management}
In \gls{satcom} networks, \gls{ml} techniques are widely used to optimize key resource utilization, including transmission power, bandwidth, and computational capacity, enabling more efficient and adaptive network operations \cite{ML6}.
\gls{ml} algorithms facilitate dynamic bandwidth allocation based on traffic demand and optimize onboard computational resources for tasks such as data compression and routing. 
Given the high cost and limited feasibility of satellite maintenance due to their altitude, \gls{ml} can also play a critical role in predictive maintenance, helping to detect potential issues before they escalate. 
These capabilities collectively enhance overall network performance and support real-time adaptation of communication parameters, ensuring seamless and reliable service even in challenging environmental conditions.

Scheduling bandwidth to improve transmission efficiency and coverage in satellite networks is a challenging problem considering the environmental dynamics. 
A \gls{madrl} approach is proposed to solve this problem, considering a \gls{geo} satellite environment in \cite{Liao2020}. 
\gls{vits} is a satellite-terrestrial integration approach discussed in literature \cite{giambene2018satellite}; however, the traffic demand for \gls{vits} is not uniformly distributed, and this initiates the requirement for flexible payload architectures. 
Hence, \cite{9448341} presents a dynamic resource management methodology leveraging \gls{drl}. 
% Authors of \cite{9399261} discuss the spectrum and infrastructure sharing for satellite-terrestrial integrated networks which is presented as a promising approach for 6G and beyond networks \cite{3GPP}. They try to solve network control and resource allocation problem using hierarchical deep actor-critic-based RL techniques in such networks. 
Further, a resource management framework for \gls{satcom} compatible with \gls{sdn}/\gls{nfv}-based management structure is studied in \cite{ML6}, which supports intercommunication with different satellite systems. 
The authors then use \gls{drl} for resource allocation in the proposed method.  
Demand-based dynamic resource allocation in satellite networks is a high computational task due to the variation and complexity of the parameters governing the problem. 
Authors in \cite{abdu2022deep} identified this as a barrier to the practical deployment of such approaches and present a methodology combining conventional optimization and \gls{dl} techniques. 
Through simulations, they show the proposed approach takes less time to optimize the parameters, resulting in less use of satellite resources.

\subsection{Training}
\begin{figure}[h]
    \centering
    \includegraphics[width=0.48\textwidth]{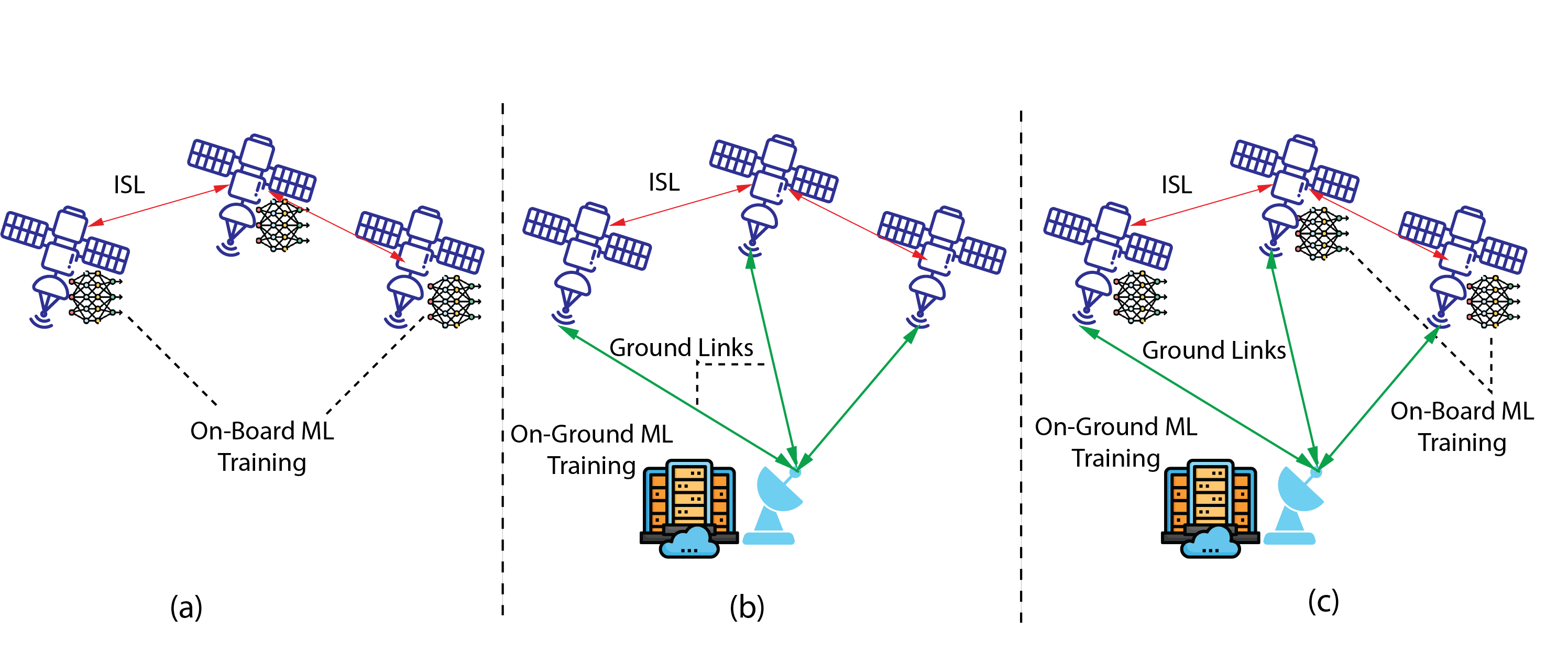}
    \caption{(a) Onboard  (b) On-ground  (c) Hybrid ML training scenarios in satellite networks}
    \label{fig:ml_training}
\end{figure}

One of the critical architectural questions needed to be addressed in adapting \gls{ml} for \gls{satcom} is training location; should training occur onboard the satellite or on the ground? 
This distinction is more than semantic, as it fundamentally affects system decision latency, adaptability, computational requirements, and overall responsiveness \cite{al2023artificial}.
As illustrated in Fig. \ref{fig:ml_training}, there can be multiple solutions to this problem, as it is directly affiliated with the \gls{qos} satellite system delivered to its ground users. 
Onboard \gls{ml} training, often conflated with online learning, entails training models directly on the satellite using data collected during its operation. 
This approach supports real-time adaptation to evolving environmental conditions, enhances autonomy, and eliminates additional latency caused by communication with ground stations \cite{ren2021interaction}. 
However, it imposes significant challenges due to the constrained power, memory, and computational capabilities of the modern compact satellites \cite{hussein2021reducing}.
Despite these limitations, recent advances in low-power AI accelerators and edge \gls{ml} chips (e.g., Intel’s Movidius and NVIDIA Jetson) are acting as key facilitators for onboard \gls{ml} deployment in satellite systems \cite{fontanesi2025artificial}.

The other option is to leverage the computational abundance of terrestrial infrastructure for \gls{ml} model training and off-board the trained model to satellite systems. 
Also referred to as on-ground training, this methodology can also benefit from the extensive data sets to develop large-scale models \cite{ortiz2023onboard, furano2020towards}. 
This approach is typically associated with offline learning, where models are pre-trained on static datasets before deployment. 
The on-ground strategy simplifies model development and version control while enabling thorough testing prior to mission deployment, leading to more robust models. 
However, it also introduces latency in dynamic decision making and lacks the adaptability to adapt to orbital anomalies and mission changes unless frequent model updates are transmitted to the satellite, which can be bandwidth and time-constrained. 
While this approach is more practical for complex and large \gls{ml} models such as \gls{llm} adaptations, its reliance on predefined training data means that unforeseen operational scenarios may lead to performance degradation.

Hybrid strategies are gaining traction to address the trade-offs between onboard and on-ground \gls{ml} training for \gls{satcom} networks. 
These architectures combine the strengths of both domains by performing online learning on the ground using near real-time or recorded satellite telemetry, then transmitting updated models or decision parameters to the satellite \cite{fontanesi2025artificial}.
Incremental learning is another hybrid approach, where a model pre-trained on the ground is fine-tuned onboard with local data, thus customizing the model with unique characteristics. 
Such configurations enable systems to adapt over time while managing satellite constraints. 
Furthermore, distributed learning also materializes solutions, allowing multiple satellites or a whole constellation to collaboratively train a global model without sharing raw data, preserving bandwidth and privacy. 
Ultimately, the choice between onboard, on-ground \gls{ml} and hybrid training must align with mission requirements, computational budgets, latency tolerance, and operational resilience, which are unique characteristics for different satellite constellations and use cases.

\subsection{Operational Resilience}

\gls{satcom} networks are prone to unexpected changes due to environmental changes and space debris \cite{ya2016mitigation, ren2021interaction}. 
Therefore, reliability and resilience of \gls{ml} for \gls{satcom} systems are critical attributes that complement performance and ensure consistent operation under those uncertain, resource-constrained, and dynamic conditions. 
Unlike performance, which measures how well an \gls{ml} agent executes its intended task, resilience incorporates and depicts the \gls{ml} agent’s ability to generalize across varying scenarios and resist failures.
In addition to the environmental uncertainties, \gls{satcom} networks pose unique challenges in operational resilience due to limited onboard computational capacity, delayed feedback loops, safety-critical operations, and partially observable high-dimensional state spaces. 
For instance, deploying \gls{drl} models directly onboard for real-time inference can be hazardous, as misinformed actions may cause irreversible satellite damage and degrade user \gls{qos}. 
To address these risks, recent studies explore model-based and offline \gls{drl} approaches, where agents learn from historical data or simulations, avoiding costly real-time experimentation \cite{xie2024computation}. 
Hardware advances, such as radiation-tolerant AI accelerators and compact edge \gls{ml} processors, now enable limited onboard training and inference, further enhancing robustness while maintaining safety margins \cite{ya2016mitigation, fontanesi2025artificial}.

Another significant concern in the robustness of \gls{ml} models in \gls{satcom} is the presence of system delays, both in communication and reward feedback. 
Traditional \gls{drl} assumes immediate feedback, which is impractical in real-world satellite networks, due to the model training location and network architecture constraints. 
Delays in observation or reward reception due to long propagation distances can lead to outdated decisions if mishandled. 
To mitigate this challenge, delay-aware \gls{mdp} frameworks and artificial training delays have been proposed to align training conditions with operational realities \cite{chen2021delay, xie2024computation}. 
Techniques such as state augmentation and temporal correlation exploitation—using models like \gls{ddpg} or Echo State Networks~(ESN), help compensate for outdated \gls{csi} and allow \gls{ml} agents to perform accurate resource allocation and power control in dynamic environments \cite{omid2024reinforcement}. 
Additionally, decentralized \gls{madrl} architectures and hierarchical clustering strategies have been explored to accelerate beam hopping and spectrum management under uncertainty \cite{zhao2023delay}. 
These innovations collectively underscore that robustness in \gls{ml}-enabled \gls{satcom} is not merely a product of model accuracy but a consequence of architecture, training realism, hardware adaptability, and resilience to systemic delay and operational variability.

\section{Performance Evaluation in Satellite Spectrum Management}
\label{sec:performance}

% In \gls{cogsat} radio environments, radio network-related measurements such as Signal to Interference plus Noise Ratio~(SINR), Received Signal Strength Indicator~(RSSI), Carrier-to-Noise Ratio~(CNR), and interference power can be categorized as foundational performance indicators of the network behaviour. 
% These metrics provide real-time insights into the radio environment, enabling intelligent decision-making for spectrum access, beam-forming, and link adaptation. 
% SINR, in particular, determines the quality and reliability of a communication link by quantifying the ratio between the desired signal and the combined interference and noise.
% Maintaining a high SINR is crucial to supporting high-order modulation schemes and achieving robust, high-throughput links. Additionally, continuously monitoring interference thresholds helps regulate transmit power and minimize spectral collisions with neighbouring systems. 
In \gls{cogsat} networks, where dynamic spectrum sharing and coexistence with terrestrial or other satellite systems are common, leveraging accurate performance evaluation methods is essential for adaptive resource management and overall system performance optimization.
This section focuses on these performance metrics and discusses their evaluation criteria. 

\subsection{Spectrum utilization}
A primary metric of overall system performance and a measure of the effectiveness of \gls{cogsat} deployment, which also offers a normalized view of how well the spectrum is utilized in shared environments.
\gls{itu} discussed metrics to measure spectrum in radio communication networks along three dimensions \cite{ITU_spectrum_effi}. 
Namely, Spectrum Utilization Factor~(SUF), Spectrum Utilization Efficiency~(SUE), and Relative Spectrum Efficiency~(RSE). 
The SUF ($U$) is defined as the product of frequency bandwidth ($B$), space ($S$),  and time ($T$), which is given in Eq. \ref{suf}. 
In which $S$ refers to the \gls{geo}metric space or area of interest, and in the context of \gls{satcom} networks, this can be a line representing a \gls{geo} or a \gls{leo} orbit. 
$T$ is the time denied to other potential users. 
\gls{itu} highlights the fact that time can be ignored in some scenarios, considering the continuity of the service. 
However, in cases such as broadcast and single-channel transmission, and in CR environments where frequency is shared, the time factor should be considered. 
\begin{equation}
    U=B.S.T
    \label{suf}
\end{equation}
The SUE is defined as a complex criterion $SUE =  \{M, U\}$, where $M$ is defined as the useful effect obtained with the aid of a network of interest. 
The \gls{itu} further simplifies this as Eq. (\ref{sue}) to the ratio between the useful effect and SUF. 
\begin{equation}
    SUE=\frac{M}{U} = \frac{B'.S'.T'}{B.S.T}
    \label{sue}
\end{equation}
In the above equation, $B', S'$, and $T'$ represent actual measurements of occupational bandwidth, coverage area and operating time, respectively. 
The RSE, which is given in the equation below, is introduced as a ratio of SUE providing the same type of service or as a ratio relative to a theoretical system. 
\begin{equation}
    RSE=\frac{SUE_a}{SUE_{std}}
    \label{rse}
\end{equation}
In Eq. (\ref{rse}), $SUE_a$ and $SUE_{std}$ are the spectrum efficiencies of actual and standard communication systems, respectively. 
In addition, spectral efficiency can be measured as the total throughput achieved per unit bandwidth. 
This reflects the overall performance improvement of the system under constrained bandwidth utilization, which is what \gls{cr} radio systems are primarily defined to achieve. 
Another approach to measure spectrum utilization is presented in Eq. (\ref{uti}), where it is presented as a function between the number of successfully allocated channels~($N_{s}$), number of channels in collision with \gls{pu} system~($N_{c}$) and the total number of channels available for sharing~($N_{T}$). 
This is more suitable for \gls{cr} systems where spectrum reuse is deployed. 
\begin{equation}
    S_{U}=\frac{N_{s} - N_{c}}{N_{T}}
    \label{uti}
\end{equation}

\subsection{Radio Network Interference}
Performance evaluation metrics, such as \gls{sinr}, \gls{snr} and \gls{inr}, are common physical layer radio network measurements that can also be leveraged to evaluate \gls{cogsat} networks.
Radio network interference is the primary factor in performance degradation in radio networks. 
The inter-network interference between \gls{pu} and \gls{su} systems and intra-network interference within the \gls{su} system should be considered in evaluating interference in the context of \gls{cogsat} systems \cite{bohai}.  
\gls{satcom} links are designed with an interference tolerance level, also called interference temperature or interference margin. 
In \cite{itu-r}, \gls{itu} recommends fixed \gls{geo}  service networks operating in frequencies below 30 GHz to design and operate their links to tolerate interference levels up to 25\% of the total system noise power when the network does not practice frequency reuse. 
This interference margin is reduced to 20\% when the networks leverage frequency reuse.

However, these average interference thresholds are defined under general communication conditions, which might not be compatible with \gls{cogsat} networks. 
In \gls{csa} \gls{cogsat} environments, a strategically formulated \glspl{ipc} is required to govern the service quality. 
\gls{ipc} regularizes the interference thresholds, and it can be deployed under two scenarios. 
Peak \gls{ipc} enforces strict interference limits for all channel states, which is suitable for protecting \gls{pu} \gls{qos}, and the average \gls{ipc} allows for higher flexibility by averaging interference over time, which is beneficial for delay-tolerant \gls{pu} applications \cite{yixuan2022underlay, musavian2009fundamental}.
When explicit \gls{ipc} constraints are unavailable, \gls{csa} performance can also be evaluated by setting a maximum tolerable performance degradation (rate or outage) for the \gls{pu}. 
It ensures that the \gls{su}'s concurrent access does not significantly degrade the \gls{pu}'s communication quality. 
This methodology requires more extensive \glspl{csi} on \gls{pu} transmit parameters, which are often challenging to obtain in practice \cite{liang2020dynamic}.
DARPA \gls{cr} implementation is a good example of peak \gls{ipc}, in which they defined and maintained 3 dB \gls{snr} degradation at a primary receiver \cite{darpa}.

\subsection{Detection and False Alarm Probabilities}
As discussed, \gls{ss} plays a key role in realizing \gls{cogsat} networks. 
\gls{osa} primarily relies on the effectiveness and reliability of \gls{ss} to detect spectrum holes in \gls{pu} transmission. 
The probability of detection, which measures the likelihood that the \gls{su} system correctly detects the presence of a \gls{pu} in a channel, is a fundamental indicator of harmful interference to the \gls{pu}s, as a higher probability of detection results in lower interference to the \gls{pu}s. 
The probability of detection depends on several factors: the sensing time, \gls{snr} of the \gls{pu} signal at the \gls{su} receiver, and the chosen detection technique (e.g., energy detection, matched filtering, or cyclostationary feature detection). 
In energy detection, for instance, detection probability is a function of the detection threshold, noise variance, and sample size. 
Achieving a higher \gls{pu} detection probability typically requires more sensing time or sensitive receivers, which in turn can impact the agility and effectiveness of \gls{cogsat} systems.

Conversely, the probability of a false alarm is the measure of the chance that an \gls{su} incorrectly detects a \gls{pu} as active when it is idle. 
A high false alarm probability leads to missed transmission opportunities for \gls{su}s, reducing the spectrum utilization efficiency; therefore, a lower probability is preferred as it implies better access to spectrum holes for \gls{su}s \cite{sharma2012satellite, cabric2004implementation}.
Technically, false alarm probability is also influenced by the detection threshold, noise uncertainty, and environmental factors such as satellite movement and Doppler shifts. 
In energy detection schemes, a lower threshold value increases detection sensitivity but also raises false alarm probability.
Therefore, careful design measures should be utilized, creating a trade-off between detection reliability and spectrum access opportunities, as the \gls{su} transmission in the \gls{pu}s channel when it is active will lead to harmful interference.

\subsection{Channel Availability}

Under \gls{dsa} paradigms, \gls{ca} refers to the channel licensed to a \gls{pu} being available for \gls{su} communications. 
Higher \gls{ca} provides more freedom for the \gls{su} channel allocation, creating less interference to the \gls{pu} system. 
In static satellite scenarios, such as fixed-beam \gls{geo} systems, \gls{ca} primarily depends on the temporal activity pattern of the \gls{pu}, which means that the channel is deemed available if the \gls{pu} is inactive.
However, in the context of \gls{leo} or \gls{meo} constellations, where satellites are inherently mobile, \gls{ca} becomes a spatio-temporal variable.
The relative motion between the primary and secondary satellite networks creates frequent changes in coverage overlap and interference regions, thereby affecting real-time \gls{ca}.
Hence, in \gls{cogsat} environments, traditional \gls{ca} estimations based solely on \gls{pu} activity become insufficient for the reliable planning of \gls{dsa}.
% In multi-orbital \gls{cogsat} networks that consist of \gls{leo} or \gls{meo} constellations, the estimation of channel availability becomes more complex due to mobility-induced topology variations. 
% Here, availability is not solely a function of PU activity but also of the relative positioning between satellites and ground users, which dynamically alters the protection zones of incumbent users.
% That also makes CA a more significant measure, reflecting the degree of freedom SUs can access channels in the CR environment.  
Executing spectrum hand-off based on sensing decisions improves \gls{ca} but introduces trade-offs concerning hand-off delays, false alarm probabilities, and data transmission durations as discussed in~\cite{lee2015channel}. 
% For \gls{cogsat} radio networks, there exists a mobility-aware estimation framework that accounts for real-time distance and movement trajectories between primary and secondary nodes, yielding more accurate channel availability predictions than static PU-activity-based models \cite{cacciapuoti2015channel}.

\subsection{Service Retainability}
Service retainability in \gls{cogsat} radio networks is a critical \gls{qos} metric that reflects the system's ability to maintain uninterrupted communication sessions once established. 
In CR operations, especially under \gls{dsa} schemes, this metric becomes increasingly important due to the opportunistic and often pre-emptive nature of spectrum access by \gls{su}s. 
A \gls{su} service may be interrupted mainly due to the arrival of a \gls{pu}, channel failure, or lack of available backup spectrum in reserved bands.
\begin{equation}
    \text{Service Retainability} = 1 - FTP
    \label{eq:sr}
\end{equation}
General form of calculating service retainability is given in Eq. \ref{eq:sr}, and the Forced Termination Probability~($FTP$) depends on multiple factors such as \gls{pu} arrival rate and effective channel assignment rate \cite{abdelgalel2024performance, tanveer2022reliability}. 
In \gls{cogsat} scenarios, service retainability will also depend on the orbital dynamics of \gls{leo} and \gls{meo} satellites in addition to the static channel characteristics, which demands refined $FTP$.

\subsection{Energy Consumption}
Satellite communication systems operate under tight energy constraints due to a lack of power generation resources and their compact nature. 
Therefore, they are forced to embrace cross-layer design, onboard processing, and energy-aware scheduling to optimize power consumption \cite{alagoz2011energy}.
Depending on the channel capacity, the power consumption of a satellite can range from 500 to 2000 W \cite{PEREZ199897}. 
Especially in the case of \gls{cogsat} networks, additional computational tasks such as intelligent decision-making, beam-forming, and associated MIMO systems require a significant amount of system power.
On the other hand, \gls{cr} capabilities can be adapted to reduce the power consumption of the \gls{satcom} systems without sacrificing performance \cite{hussein2021reducing}. 
Therefore, the amount of successfully transmitted data units per unit of energy consumed can be a \gls{kpi} for the overall \gls{cogsat} system, reflecting on power utilization. 

Energy-aware \gls{cr} functionalities allow satellites to adaptively select modulation, coding, and power parameters based on environmental feedback, thereby minimizing redundant retransmissions and idle power dissipation. 
Metrics such as bits-per-Joule and energy-per-bit have been proposed to quantify these optimizations, ensuring that throughput remains efficient relative to the energy consumed per unit transmission \cite{hussein2021reducing, alagoz2011energy}. 
Similar to mobile communication devices, power utilization is a primary metric in communication satellites and should be monitored regularly when operating under cognitive conditions. 
In addition, integration of \gls{ml} and \gls{ai} demands a significant amount of power from compact satellite systems, which primarily depend on solar power. 
This further highlights the importance of power consumption as a performance indicator in \gls{cogsat} networks.

\subsection{Latency, Delay and Jitter}
Due to the signal traveling distance and associated \gls{ss} and processing, \gls{cogsat} networks are affected by additional delay components compared to general \gls{satcom} networks. 
Delay, the total time a data packet or signal takes to travel from the source to the destination, can be decomposed into several components such as transmission, propagation, process or decision and queuing delay. 
Latency mainly refers to the \gls{rtt} or the reaction time. 
Jitter is the variation or inconsistency in the delay experienced by consecutive packets. 
Although propagation delay and jitter are primary parameters defining \gls{qos}, in \gls{cogsat} systems, the delay associated with the process and queuing delay are of interest, as that is mostly affected by the \gls{cr} functions. 
% They are primarily associated with additional computation and decision-making processes that cognitive radio methods introduce. 

Processing latency refers to the time a \gls{cogsat} system takes to sense the spectrum environment, process the contextual information, and make an intelligent transmission decision—such as channel selection, power adaptation, or interference management. 
In CR environments, this translates into a critical \gls{kpi} that directly impacts the system’s responsiveness and performance to real-time spectrum variations. 
Especially in \gls{osa} setups, performance latency reduces the \gls{su} transmission time, thus impacting the system \gls{qos}.
Queuing in communication systems has been studied extensively in the literature, and advanced queuing algorithms have been proposed. 
However, \gls{cogsat} networks require novel queuing algorithms due to the \gls{ss} and intelligent decision-making processes incorporated in \gls{cogsat} networks. 
The goal of these performance metrics is to push \gls{cogsat} systems to bridge the gap between them and general \gls{satcom} networks, thus providing seamless transmission between the two networks.

\subsection{Communication Overhead}

In \gls{cogsat} networks, signal overhead refers to the additional signaling required to facilitate \gls{dsm}, coordination among satellites, and real-time \gls{ss}. 
It quantifies the proportion of communication resources, such as bandwidth, power, and time consumed for control signaling rather than payload data \cite{hui2025review}.
The overflowing coordination and control messages, which add up to signal overhead, can lead to bandwidth and latency degradation in a communication system. 
In \gls{cogsat} systems, especially those employing techniques like \gls{osa}, signaling overhead arises from \gls{ss} reports, channel allocation decisions, hand-off signaling, and inter-satellite coordination.
Additionally, due to the dynamic learning, adaptation, and optimization algorithms, an additional overhead/ signaling component is introduced in \gls{cr} systems. 
These control signals are prioritized over the general traffic due to their importance to the system's operation, thus adding a latency factor for the low-prioritized traffic. 
Therefore, the signaling overhead to payload ratio is a metric that reflects the efficiency of the communication protocols deployed in a \gls{cogsat} system. Minimizing this ratio is crucial in bandwidth constrained and latency sensitive \gls{cogsat} communication systems, thus highlighting its importance as a key performance metric in such systems.

\section{Challenges and Future Directions}
\label{sec:challanges}

\subsection{Regulatory Challenges}

\paragraph{Global Consensus} Implementation of spectrum sharing and trading in \gls{cogsat} communications faces significant challenges due to inadequate regulatory frameworks. 
Without generalized and agreed standards on spectrum sharing, satellite operators have to make individual decisions, thus making the process complex and unmanageable, due to the number of satellite network operators, regional and national spectrum management, alongside security regulations.  
Therefore, coordination between national and international authorities is paramount in realizing \gls{cogsat} networks, as terrestrial spectrum management is facilitated at the national level, while international cooperation is essential for satellite spectrum management. 
\gls{itu}, being the recognized body in managing global spectrum regulatory requirements, coming up with a consensus to realize a global \gls{cogsat} framework mainly relies on them. 
Clear definitions of \gls{eirp} and out-of-band interference thresholds should be defined and agreed \cite{sharma2013satellite}. 
Additionally, the secondary dynamic access to dedicated bands for the government and military requires careful regulatory attention, particularly because of the associated security risks and emergency availabilities \cite{pedram2023role}.

\paragraph{Compatibility and Spectrum Ownership} Similar to the involvement of policymakers in building a platform for \gls{cogsat} communication networks, satellite operators and equipment manufacturers have a pivotal role to play in realizing successful \gls{cogsat} networks. Equipment manufacturers have to explore methodologies to develop affordable communication equipment compatible with both terrestrial and satellite networks. 
Furthermore, business models have to be developed to share the spectrum ownership between the networks \cite{zhang2022spectrum, hassan2021survey}. 
Below are several possible modes of spectrum ownership
\begin{itemize}
    \item Temporarily transfer of usage rights to another entity on a short or medium-term basis, including the full transfer of associated rights and responsibilities. 
    \item Temporarily lease on a short-term basis, to be used based on the traffic demand. The primary holders retain their rights and obligations to the shared spectrum. 
    \item Spectrum trading, where the primary holders may also retain their rights for the traded spectrum. 
    \item Spectrum pooling, which is implemented as either pure pooling or a hybrid approach (e.g., combining fixed bands with a shared pool).
\end{itemize}
Collaboration across all stakeholders—regulators, industries, and researchers is key to overcoming these regulatory challenges and ensuring the successful deployment of \gls{cogsat} systems.

\paragraph{Enhanced Protocols} The unavailability of communication protocols for \gls{cogsat} is another challenge that standardization bodies need to address, as their absence can lead to significant interoperability challenges between the communication nodes in a \gls{cogsat} network. 
Further, these protocols should have seamless integration capabilities with the existing globally recognized communication protocols, thus mitigating the potential inefficiencies or conflicts in network deployments. 
The absence of standard protocols for \gls{cogsat} systems has hindered the trust in \gls{cogsat} networks as a whole, among the satellite network operators and equipment manufacturers, creating an obstacle in realizing \gls{cogsat} networks. 
Data privacy and security of the \gls{cogsat} network are another aspect lack in standardization. 
Particularly when \gls{pu} and \gls{su} operate in shared spectrum environments, sensitive information can be exposed through unauthorized access \cite{RSMA5}.
Additionally, concerns related to national security and integrity can arise due to the potential use of \gls{cogsat}s for unauthorized surveillance and eavesdropping, which also emphasizes the requirement for advanced, regularized security measures \cite{pedram2023role}.

\subsection{Architectural Challenges}

\paragraph{Cooperative Networks} \gls{cr} deployments in \gls{stn}s and multi-orbital satellite networks demand unified network architectures and seamless operation between the networks \cite{janu2022machine, yao2023green}. 
The relative motion between non-\gls{geo} satellites and users generates complex dynamics. 
Therefore, sophisticated spectrum sharing \gls{cogsat} techniques should be utilized to predict and adapt the radio conditions in such environments, considering the associated spatial, temporal, and spectral parameters. 
These \gls{cr} methodologies often assume a cooperative architecture with geographical location and frequency parameters of the users shared between the network of interest \cite{9628097, ruan2018performance, bohai}.
In addition, this coordination should account for the propagation characteristics and data processing delays in making real-time decisions. 
Therefore, refined cooperative network paradigms should be developed to cater to these unique requirements. 

\paragraph{Latency and Delay} These are inherent challenges in \gls{satcom}, which can also affect real-time decision making and the responsiveness of \gls{cogsat} networks \cite{xie2024computation}. 
In \gls{cogsat} systems, the radio transmission parameters adjustments, frequency reuse, and network routing decisions have to be swift and actionable, considering the dynamic conditions to minimize interference. 
However, the considerable propagation delays between the satellite and ground station, specifically in \gls{geo} communication, or between the different satellites in multi-orbital cooperative networks, can introduce complications in coordinating these decisions.
For instance,  \glspl{leo} demand rapid handovers and real-time adjustments for successful deployments of CR techniques. 
These operations are sensitive even to minor delays, which can result in \gls{qos} degradations in the \gls{cogsat} deployments. 
Therefore, effective latency and delay management strategies, such as predictive algorithms, advanced caching, and data prioritization, are essential for an efficient \gls{cogsat} network operation to maintain the \gls{qos} levels \cite{omid2024reinforcement, zhao2023delay}.

\paragraph{Scalability} Large-scale communication networks like satellite and mobile networks face rapid expansions, considering the growing global demand for connectivity. 
Therefore, \gls{cogsat} networks must be designed to cater for rapid horizontal and vertical network expansions without compromising performance, reliability, or efficiency.
Current mega constellations contain several thousand \glspl{leo}; thus the \gls{cogsat} techniques should be capable of handling seamless communication, data processing, and spectrum management across a vast, distributed system. 
This demands advanced algorithms for dynamic satellite resource allocation under \gls{cr} settings to maintain the \gls{qos} levels even under fluctuating conditions. 
Additionally, the \gls{cogsat} methodologies must efficiently manage handovers, particularly in \gls{leo} and terrestrial coexisting CR networks, where both networks experience rapid handovers due to the user dynamics, further necessitating frequent communication to different ground stations or other satellites. 
Consequently, effective, scalable \gls{cogsat} solutions must be developed to meet global demand and compatible with diverse applications, while maintaining the optimal performance levels in a growing network infrastructure.

\paragraph{Energy Efficiency} Power and computation are limited resources in satellite networks, as increasing them is associated with the manufacturing and launching costs, which can escalate the capital expenditure \cite{hussein2021reducing}.
Modern mega-satellite constellations have successfully reduced the weight of a \gls{leo} satellite to reduce the launching cost (the latest version of Starlink \gls{leo} only weighs 260 kg \cite{starlink_weight}), thus limiting the onboard power and processing capabilities. 
However, this poses a significant challenge in realizing \gls{cogsat} networks, as they require substantial computational resources to perform real-time \gls{ss}, data analysis, decision-making, and dynamic network adjustments \cite{8786872, ortiz2022supervised}. 
Apart from these functions, setups such as dual-\gls{cogsat} networks demand frequent communication and coordination between neighboring satellites and ground network operation centers.  
This requirement is further elevated in hybrid \gls{cogsat} networks, where rapid coordination between terrestrial networks is essential.
To address these limitations, \gls{cogsat} systems must employ energy-efficient algorithms and optimize resource allocations. Therefore, satellites can potentially offload some processing tasks to ground stations or more capable satellites, relaxing the strain on low-powered satellites. 
% Deploying advanced cognitive functions with limited onboard computational and power resources is a challenge that needs to be addressed to realize the successful deployment and operation of \gls{cogsat} networks.

\paragraph{Security and Privacy} These are two areas of paramount importance in modern networks, and they are further emphasized in \gls{satcom}, considering the global access and coverage spanning geographical boundaries \cite{pedram2023role}. 
The spectrum sharing and sensing techniques used in the cooperative network architectures of \gls{cogsat} networks are prone to vulnerabilities due to information sharing between the associated networks. 
Unauthorized access, eavesdropping, data interception, jamming, spoofing, and malicious use of spectrum resources are a few key vulnerabilities \gls{cogsat} networks might experience.
In Hybrid \gls{cogsat} networks, terrestrial network integration adds another layer of complexity due to the secure communication requirements between heterogeneous network interfaces. 
Similarly, Dual \gls{cogsat} networks demand secure cooperation between the satellite networks amid the dynamics of the environment. 
Facilitating these security requirements through advanced encryption and authentication processes can be challenging due to the limited computational resources in satellites. 
Therefore, \gls{cogsat} networks should be equipped with a simple but robust, multi-layered security architecture. 
It should include strong encryption, authentication protocols, intrusion detection systems, and secure key management mechanisms compatible with the employed \gls{cr} techniques \cite{perera2024survey,  chen2022blockchain}. 
Additional privacy-preserving techniques must be deployed in the \gls{cogsat} setup to protect data integrity, considering the personal and confidential information communicated through modern broadband networks. 
Further, the \gls{cogsat} network architecture should be resilient in detecting and mitigating in real-time, amid the dynamic network parameter changes and the decentralized nature of the \gls{cogsat} operations. 

\paragraph{Adapting \gls{sdn} and \gls{nfv}} Empowering \gls{satcom} networks with these two technologies offers substantial benefits in realizing \gls{cogsat} networks as they offer flexibility and programmability to network setups \cite{li2016using}.  
\gls{cr} networks require real-time control of network resources to manage the dynamic behaviors, which \gls{sdn} can efficiently handle through its inherently flexible and centralized control mechanism.
Programmability and network-wide optimization are additional features that \gls{cogsat} networks can benefit from the\gls{sdn} architecture, as it enables complex network policy implementations and offers agility to maintain the network's \gls{qos} through seamless coordination \cite{qi2022sdn, jiang2023software}. 
\gls{nfv} enables the virtualization of \gls{cr} functions, allowing these functions to run on general-purpose hardware rather than specialized, dedicated devices \cite{LP9,maity2024traffic}. 
This increases the flexibility and scalability of deploying \gls{cr} functionalities across the \gls{satcom} network. 
One of the key challenges satellite networks face is the lack of flexibility to deliver new services with the existing hardware, as hardware upgrades in orbiting satellites are practically impossible. 
\gls{nfv} provides a solution to that problem, thus enabling the deployment of upgraded cognitive solutions with the same hardware. 
However, these technologies are yet to be fully realized in \gls{satcom} networks, posing a significant challenge in realizing \gls{cogsat} networks.

\subsection{Machine Learning Implementations}

\paragraph{Heterogeneity} Integrating \gls{ml} algorithms into \gls{cogsat} networks presents substantial challenges, particularly when interfacing with existing systems. 
\gls{ml} models can be leveraged for tasks such as \gls{dsm}, resource optimization, and interference mitigation in \gls{cogsat} systems (refer Table \ref{tab:ml_cr}). 
However, these cognitive capabilities empowered through \gls{ml} should be compatible with existing satellite and terrestrial networks that operate on fixed frequency allocations and standardized communication protocols.
The heterogeneity of these systems creates significant barriers to interoperability, as almost all the existing networks lack the flexibility to accommodate adaptive decision-making processes demanded by \gls{cogsat} networks.
Furthermore, the seamless integration of \gls{ml}-based cognitive functionalities requires extensive modifications to existing network architectures in addition to general \gls{cr} methods, thus necessitating the deployment of middleware solutions or protocol converters specifically catered to ensure \gls{ml} model compatibility between \gls{cogsat} and general networks.
The challenge is further complicated by the need to maintain backwards compatibility, which can present significant limitations to the extent to which \gls{ml}-driven \gls{cr} innovations can be fully realized.
Ensuring smooth operation across \gls{cogsat} and legacy systems demands robust and interoperable \gls{ml} frameworks capable of managing the complex interactions between these networks while providing accurate network predictions amid the complex dynamics of the \gls{satcom} environment.

\paragraph{Communication Overhead and Security}
% In addition to integration challenges with existing networks, \gls{cogsat} networks must address the significant communication overheads and security and privacy concerns that arise from \gls{ml} implementation.
\gls{cogsat} networks demand continuous data exchange between satellites, ground stations, and other network elements to maintain synchronized \gls{cr} functions. 
These data exchanges, essential for \gls{ml} model updates, coordination, and real-time decision-making in \gls{cogsat}s, introduce considerable communication overhead, particularly given the limited bandwidth and high latency characteristic of \gls{satcom} links.
The requirement to balance data throughput amid energy efficiency and processing constraints exacerbates this challenge.
Therefore, advanced \gls{ml} algorithms capable of reducing the communication overhead need to be developed.
In addition, data transmission approaches should be cautiously chosen, communicating only necessary updates after the initial handshake. 
Moreover, \gls{ml} integration into \gls{cogsat} networks heightens the importance of security and privacy.
% Satellites networks are vulnerable to a range of cyber threats, including adversarial attacks, data breaches, and signal jamming.
% This situation is further degraded in \gls{cogsat} networks due to shared spectrum use with advanced distributed sensing capabilities, thus exposing the users and the network to intruders at unprecedented levels. 
\gls{ml} models in \gls{cogsat} networks often operate in the control plane, therefore, to protect the satellite networks and \gls{ml} model's performance, advanced cryptographic techniques, such as homomorphic encryption and secure multi-party computation, should be implemented within the \gls{ml} algorithms to safeguard data during transmission and processing.
Additionally, privacy-preserving approaches such as differential privacy should be incorporated into \gls{cogsat} systems to protect the integrity of sensitive information, even as data is shared and processed across the network. 
Addressing these challenges is vital in realizing \gls{ml}-driven \gls{cogsat} networks to ensure they operate in a secure and efficient manner.

\paragraph{Data Scarcity}
The effectiveness of a \gls{ml} model relies heavily on the quality and quantity of data available for training and validation.
However, in the context of \gls{satcom} networks, obtaining such data representing the network's diverse and dynamic nature has practical limitations. 
Satellites operate across different orbits, covering vast geographical areas with separate frequencies, leading to highly diverse conditions that are difficult to encapsulate comprehensively in a dataset. 
Moreover, labeled data specific to \gls{satcom} scenarios, such as spectrum usage patterns, interference levels, and environmental effects, is often scarce or unavailable.
In addition, \gls{cogsat} networks have not been realized, thus adding the requirement of converting/\gls{rem}odeling data captured in current satellite deployments. 
This scarcity limits the effective learning abilities of \gls{ml} models. 
Additionally, the quality of the data is often compromised by incomplete information due to unsynchronized, antagonistic data collection methods, which can degrade \gls{ml} model performance. 
Therefore, to address these challenges, there is a need for advanced data collection methods specifically catered to satellite environments, guaranteeing both quality and quantity. 
As an alternative approach, synthetic data generation methods can be explored, considering practical limitations.

% Another challenge in leveraging \gls{ml} for \gls{cogsat} networks lies in computational and power resource constraints. 
% Modern \glspl{leo} have limited onboard processing power, memory, and storage, which restricts their ability to run complex \gls{ml} algorithms in real time. 
% Practical issues in upgrading the SATs with additional computational power add another layer of complexity to the problem. 
% However, \gls{ml} implementation is essential in predicting radio parameter configurations amid complex environmental dynamics to achieve optimal performance in \gls{cogsat} networks. 
% Furthermore, the need for energy efficiency compounds the problem, as SATs operate under stringent power budgets due to their dependency on solar energy and limited battery capacity.
% Therefore, running resource-intensive \gls{ml} models can quickly deplete available energy, impacting critical orbital maneuvers, communication, and payload operations.  
% As a result, there is a pressing need to develop lightweight, energy-efficient \gls{ml} algorithms that can operate within these constraints while still delivering the performance required for cognitive decision-making.
% \gls{ml} paradigms such as transfer learning, distributed learning, and model comparisons in addition to the computational resource optimization approach in edge computing and task prioritization methods should be employed to balance the inherent computation and power limitations in \gls{cogsat} networks. 

\paragraph{Generalization}
% Generalized model development is a critical challenge in realizing \gls{ml} deployments within \gls{cogsat} networks.
For \gls{ml} models to be effective in \gls{cogsat} environments, they must generalize profoundly across various network and communication conditions. 
However, the highly dynamic and diverse nature of \gls{satcom} makes this difficult.
\gls{cogsat} networks are multi-orbital and must operate in varying spectral conditions and under fluctuating network conditions, often with limited prior data, as highlighted above.
An \gls{ml} model trained on data from a specific scenario may not perform accurately under new, unseen situations, which might lead to poor decision-making and reduced \gls{qos}.
This lack of generalization can be particularly problematic when dealing with rare or ext\gls{rem}e events, such as unexpected interference or sudden changes in spectrum availability.
To enhance \gls{ml} model generalization, it is essential to develop robust training strategies that expose the model to a wide variety of conditions, further highlighting the challenge of data scarcity.
Methodologies such as transfer learning and behavior cloning can improve the \gls{ml} model's adaptation to new environments, and implement continuous learning approaches where the model evolves as it encounters new data.
% Therefore, ensuring that \gls{ml} models can generalise across the broad spectrum of challenges in \gls{cogsat} networks is paramount for achieving reliable and adaptable network performance, thus delivering \gls{qos}-guaranteed service to the users.

\paragraph{Scalability}
% In realizing \gls{ml} for \gls{cogsat} networks, the scalability of the \gls{ml} models is of paramount importance considering the rapid expansions and complexity of the modern SAT constellations. 
In parallel to satellite constellation expansions, especially with the advent of mega \gls{leo} constellations consisting of thousands of satellites, the \gls{ml} models must scale effectively to handle the growing volume of data and the increasing number of nodes.
Thus the \gls{ml} models in \gls{cogsat} networks should have the capability to handle the increasing data inputs while ensuring the ability to make real-time decisions across a distributed network without compromising performance.
The distributed network architecture of \gls{cogsat} networks, where satellites collaborate across multiple orbits and coordinate with ground stations, further complicates scalability in such systems.
centralized \gls{ml} approaches may struggle to cope with the sheer scale, which further necessitates the decentralized learning methods for \gls{cogsat} systems that distribute the learning process across multiple satellites or ground stations.
As the categorization highlights in Table \ref{tab:ml_cr}, there exists a clear gap in the literature in distributed learning for \gls{cogsat}.
% Computational resource limitations and higher communication overheads are also associated challenges with \gls{ml} scalability, therefore advanced \gls{ml} models should be developed to mitigate these performance bottlenecks leading to \gls{cogsat} network performance degradations. 
Therefore, developing \gls{ml} models that can scale efficiently while maintaining robustness, accuracy, and responsiveness is essential to realize \gls{cogsat} networks on a large scale.

\section{Conclusions}
\label{conclusion}

The traditional exclusive satellite frequency allocation approach is leading to spectrum scarcity, which creates barriers for new \acrfull{satcom} operators while limiting the capabilities of the existing service providers. 
Intelligent spectrum management approaches enabled by \acrfull{cogsat} provide feasible solutions to this upcoming issue, and \acrfull{ai}/ \acrfull{ml} with superior decision-making and classification capabilities is identified as a key enabler of \acrfull{cogsat} systems. 
In this paper, we highlighted the unique characteristics of existing satellite systems and their roles in facilitating global communication through inter-satellite and satellite-terrestrial integrations. 
Furthermore, this paper extensively evaluated \acrfull{cr}-enabled dynamic spectrum management approaches for \gls{satcom} in the context of \acrfull{osa} and \acrfull{csa} and discusses the state-of-the-art \gls{ml} approaches leveraged in \acrfull{ss}, allocation, interference mitigation, and resource management for \gls{cogsat} networks. 
However, the deployment of these \gls{cr} methodologies in satellite networks enabling \gls{cogsat} remains a complex challenge, due to the barriers caused by regulatory bodies and the limitations in network architecture and standardization. 
Moreover, leveraging \gls{ml} in \gls{cogsat} has challenges due to the computational and power constraints in modern satellite systems. 
This paper discusses these issues and details \gls{ml} as a promising solution for scalable, efficient, and secure enabler of \gls{cogsat} systems toward sustainable future satellite networks as a solution for global connectivity.

\section*{ACKNOWLEDGMENT}
The authors acknowledge the guidance from Dr. Z. Krusevac for this work, who is with the Defence Science and Technology Group (DSTG), Australia.

\bibliographystyle{IEEEtran}
\bibliography{IEEEabrv, cogsat_corrected}

\end{document}